\documentclass[journal]{IEEEtran}

\usepackage{amsmath,amssymb,amsfonts,amsthm,empheq}
\usepackage{graphicx} 
\usepackage{color}
\usepackage{multirow}
\usepackage{stfloats}
\usepackage{mathrsfs}
\usepackage{array}
\usepackage{dsfont}

\newtheorem{theorem}{Theorem}
\newtheorem{lemma}{Lemma}

\newcommand{\SNR}{\mbox{\textsf{SNR}}}
\newcommand{\Er}{\mbox{\textsf{Er}}}
\newcommand{\beq}{\begin{equation}}
\newcommand{\eeq}{\end{equation}}
\newcommand{\bea}{\begin{eqnarray}}
\newcommand{\eea}{\end{eqnarray}}
\newcommand{\BO}{\mathcal{B}_{\Omega}} 
\newcommand{\BBO}{\underline{\mathcal{B}}_{\Omega}} 
\newcommand{\BOP}{\mathcal{B}_{\Omega}^{'}}
\newcommand{\BBOP}{\bar{\mathcal{B}}_{\Omega}^{'}}
\newcommand{\BOPP}{\mathcal{B}_{\Omega}^{''}} 
\newcommand{\HBOP}{\hat{\mathcal{B}}_{\Omega}^{'}}
\newcommand{\f}{\mathsf{f}}
\begin{document}

\sloppy


\title{Information Without Rolling Dice}


\author{
	Taehyung~J.~Lim and Massimo~Franceschetti,~\IEEEmembership{Senior Member, IEEE}  
	\thanks{Taehyung~J.~Lim and Massimo Franceschetti are with the Information Theory and Applications Center (ITA) of the California Institute of Telecommunications and Information Technologies (CALIT2), Department of Electrical and Computer Engineering, University of California, San Diego CA, 92093, USA. Email: taehyung.lim@hotmail.com, massimo@ece.ucsd.edu.}
\thanks{This work is partially supported by the National Science Foundation
award CCF-1423648. }
}

\maketitle


\begin{abstract}
The deterministic notions of capacity and entropy are studied in the context of communication and storage of information using square-integrable, bandlimited signals subject to perturbation. The $(\epsilon,\delta)$-capacity, that extends the Kolmogorov $\epsilon$-capacity to packing sets of overlap at most $\delta$, is introduced and compared to the Shannon capacity.  The functional form of the results indicates  that in both Kolmogorov and Shannon's settings, capacity and entropy  grow linearly with the number of degrees of freedom, but only logarithmically with the signal to noise ratio. This basic insight transcends the details of the stochastic or deterministic description of the information-theoretic model.	
For $\delta=0$, the analysis leads to new bounds on the Kolmogorov $\epsilon$-capacity, and to a tight asymptotic expression of the Kolmogorov  $\epsilon$-entropy of bandlimited signals. A deterministic notion of error exponent is introduced. Applications of the theory are briefly discussed.

\end{abstract}



\section{Introduction}

\PARstart{C}{l}aude Shannon  introduced the notions of  capacity and  entropy in the context of communication in 1948~\cite{shannon1948}, and with them he ignited a technological revolution. His work instantly became a classic and it is   today the pillar of modern digital technologies. On the other side of the globe, the great Soviet mathematician Andrei Kolmogorov was acquainted with  Shannon's work in the early 1950s and immediately recognized that ``\emph{his mathematical intuition is remarkably precise.}'' His notions of  $\epsilon$-entropy and $\epsilon$-capacity~\cite{kolmogorov1956, kolmogorov1961} were certainly influenced by Shannon's work.   The $\epsilon$-capacity has the same operational interpretation of Shannon's in terms of the limit for the amount of information that can be transmitted under perturbation, but it was developed in the purely deterministic setting of functional approximation. 
On the other hand, the $\epsilon$-entropy corresponds to the amount of information required to represent any function of a given class  within $\epsilon$ accuracy, while the Shannon entropy corresponds to the average amount of information required to represent any stochastic process of a given class, quantized at level $\epsilon$. 
Kolmogorov's   interest in approximation theory dated back to at least the nineteen-thirties, when he introduced the concept of $N$-width to characterize the ``massiveness''  or effective dimensionality of an infinite-dimensional functional space~\cite{kolmogorov}. This interest also eventually led him to the solution in the late nineteen-fifties, together with his student Arnold, of Hilbert's thirteenth problem~\cite{kolmogorov1957}. 

Even though they shared the goal of mathematically describing the limits of communication and storage of information, Shannon and Kolmogorov's approaches to information theory  have evolved separately. Shannon's  theory flourished in the context of communication, while Kolmogorov's work impacted mostly mathematical analysis. Connections between their definitions of entropy have been pointed out in~\cite{donoho}, and we discussed the relationship between capacities in our previous work~\cite{allerton1}. The related concept of complexity and its relation to algorithmic information theory has been treated extensively~\cite{coverbook,gray1989}. 
Kolmogorov devoted  his presentation at the 1956 International Symposium on Information Theory~\cite{kolmogorov1956isit}, and Appendix~II of his work with Tikhomirov~\cite{kolmogorov1961} to explore the relationship with the probabilistic theory of information developed in the West, but limited the discussion ``\emph{at the level of analogy and parallelism.}''  This is  not surprising, given the state of affairs of the mathematics of functional approximation in the nineteen-fifties --- at the time the theory of  spectral decomposition of time-frequency limiting operators, needed for a rigorous treatment of continuous waveform channels,  had yet to be developed by Landau,  Pollack and Slepian~\cite{Slepian1976, slepian1983}. 


Renewed interest in deterministic models of information has recently been raised in the context of networked control theory~\cite{matveev, nair2013}, and in the context of electromagnetic wave theory~\cite{massimo2014,FraMigMin2011,FraMigMin2009}. Motivated by these applications, in this paper we define the number of degrees of freedom, or effective dimensionality, of the space of bandlimited functions in terms of $N$-width, and study capacity and entropy in Kolmogorov's deterministic setting.  We also extend Kolmogorov's capacity to packing sets of non-zero overlap, which allows a more detailed comparison with 
Shannon's work.

%

\subsection{Capacity and packing}
Shannon's capacity is  closely related to  the problem of geometric packing ``billiard balls'' in high-dimensional space.  Roughly speaking, each transmitted signal, represented by the coefficients of an orthonormal basis expansion, corresponds to a point in the space, and  balls centered at the transmitted points represent the probability density of the uncertainty of the observation performed at the receiver.
A certain amount of overlap between  the balls  is allowed to construct dense  packings corresponding to codebooks of high capacity, as long as   the overlap does not include typical noise concentration regions, and  this allows to achieve reliable communication with vanishing probability of error. 
The more stringent requirement of communication with probability of error equal to zero leads to the notion of zero-error capacity~\cite{shannon1956}, which depends only on the region of uncertainty of the observation, and not on its probabilistic distribution, and it can be expressed as the supremum of a deterministic information functional~\cite{nair2013}.

Similarly, in Kolmogorov's    deterministic setting communication   between   a transmitter and a   receiver occurs without error,  balls of fixed radius $\epsilon$ representing the uncertainty introduced by the noise about each transmitted signal are not allowed to overlap, and his notion of $2\epsilon$-capacity corresponds to the Shannon zero-error capacity of  the $\epsilon$-bounded  noise channel. 

In order to represent a vanishing-error  in a deterministic setting,  we allow a certain amount of overlap between the $\epsilon$-balls. In our setting, a codebook is composed by a subset of waveforms in the space, each corresponding to a given message. A transmitter can  select  any one of these signals,  that is observed at the receiver  with perturbation at most $\epsilon$. If signals in the codebook are at  distance less than $2 \epsilon$ of each other, a decoding error may occur  due to the overlap region between the corresponding $\epsilon$-balls.  The total volume of the error region, normalized by the total   volume of the $\epsilon$-balls in the codebook, represents a  measure of the fraction of space where the received signal may fall and result in a communication error. The $(\epsilon,\delta)$-capacity is then defined as the logarithm base two of the largest number of signals that can be placed in a codebook having  a normalized error region of size at most $\delta$. We provide upper and lower bounds on this quantity, when communication occurs using bandlimited, square-integrable signals, and introduce a natural notion of deterministic error exponent associated to it, that depends only on the communication rate,  on $\epsilon$,  on the signals' bandwidth,  and on the energy constraint. Our bounds become tight for high values of the signal to noise ratio, and their functional form   indicates that capacity  grows linearly with the number of degrees of freedom, but only logarithmically with the signal to noise ratio. This  was  Shannon's original insight, revisited here in a deterministic setting.

For $\delta=0$ our notion of capacity reduces to the Kolmogorov $2\epsilon$-capacity, and we provide new bounds on this quantity.
By comparing the lower bound for $\delta>0$ and the upper bound for $\delta=0$, we also show that  a strict inequality holds between the corresponding values of capacity if the signal to noise ratio is sufficiently large. The analogous result in a probabilistic setting is  that the Shannon capacity of the uniform noise channel  is strictly greater than the corresponding zero-error capacity.


\subsection{Entropy and covering}
Shannon's entropy is closely related to the geometric problem of covering a high-dimensional space with balls of given radius. Roughly speaking, each  source signal, modeled as a stochastic process, corresponds to a random point in the space, and by quantizing  all coordinates of the space at a given   resolution, Shannon's entropy corresponds to the number of bits needed on average to represent the quantized signal. Thus, the entropy  depends on both the probability distribution of the process, and  the  quantization step along the coordinates of the space.  A quantizer, however, does not need to act uniformly on each coordinate, and can be more generally  viewed   as a discrete set of balls covering the   space. The source signal is  represented by the closest center of a ball covering it, and the distance to the center of the ball represents the distortion measure associated to this representation. In this setting, Shannon's rate distortion function provides  the minimum number of bits that must be specified per unit time to represent the source process with a given average distortion.

%
%
%
%
In Kolmogorov's   deterministic setting, the $\epsilon$-entropy is the logarithm of the minimum number of balls of radius $\epsilon$ needed to cover the whole space and, when taken per unit time, it corresponds to the Shannon  rate-distortion function, as it also represents  the minimum number of bits that must be specified per unit time to represent any source  signal with distortion at most $\epsilon$.
We provide a tight expression for this quantity, when sources are  bandlimited, square-integrable signals.
The functional form of our result shows that the $\epsilon$-entropy grows linearly with the number of degrees of freedom and logarithmically with the ratio of the norm of the signal to the norm of the distortion. Once again, this was Shannon's key insight that remains invariant when subject to a deterministic formulation.

The \emph{leitmotiv} of the paper is the comparison between deterministic and stochastic approaches to information theory, and the presentation is organized as follows:
In Section II we informally describe our results, in section III we present our model rigorously, provide some definitions, recall results in the literature that are useful for our derivations, and present our technical approach. Section IV briefly discusses applications. Section~V provides  precise mathematical statements of our results, along with their proofs.  A discussion of previous results and the computation of the error exponent in the deterministic setting appear in the Appendixes.

%


\section{Description of the results}
We begin with an informal description of our  results, that is placed on rigorous grounds in subsequent sections.
\subsection{Capacity}
We consider one-dimensional, real, scalar waveforms of a single scalar variable  and supported over an angular frequency interval $[-\Omega,\Omega]$. We assume that waveforms are square-integrable,  and satisfy  the energy constraint
\beq
	\int_{-\infty}^{\infty} f^2(t) dt  \leq E. 
\label{eq:ecn}
\eeq
These bandlimited waveforms have unbounded time support, but are observed over a finite interval $[-T/2,T/2]$. In this way, and in a sense to be made precise below,   any signal  can be expanded  in terms of  a suitable set of  basis functions, orthonormal over the real line,  and for $T$ large enough it can be seen as a point in a space of essentially 
\beq
	N_0 = \Omega T /\pi
\eeq 
dimensions, corresponding to the number of degrees of freedom of the waveform, and of radius $\sqrt{E}$. 

To introduce the notion of capacity, we consider an uncertainty sphere of radius $\epsilon$ centered at each signal point, representing the energy of the noise that is added to the observed waveform. In this model, due to Kolmogorov, the  signal to noise ratio is
\beq
	\SNR_K = E/\epsilon^2.
 \eeq
 
A codebook is composed by a subset of waveforms in the space, each corresponding to a given message. A transmitter can  select any one of these signals,  that is observed at the receiver  with perturbation at most $\epsilon$. By choosing signals in the codebook to be at at distance at least $2 \epsilon$ of each other, the receiver can decode the message without error. The   $2\epsilon$-capacity  is the logarithm base two of the   maximum number $M_{2\epsilon}(E)$ of distinguishable signals in the space. This geometrically corresponds to the maximum number of  disjoint balls of radius $\epsilon$ with their centers situated inside the signals' space and it is given by
 \beq
 	C_{2 \epsilon} = \log M_{2\epsilon}(E) \;\;\; \mbox{bits}.
 \eeq
We also define the capacity per unit time
 \beq
	\bar{C}_{2 \epsilon} = \lim_{T \rightarrow \infty} \frac{\log M_{2\epsilon}(E)}{T} \;\;\; \mbox{bits per second}.\
\label{cdef}
 \eeq

A similar Gaussian stochastic model, due to Shannon, considers  bandlimited signals in a space of essentially $N_0$ dimensions, subject to an energy constraint over the interval $[-T/2,T/2]$ that scales linearly with the number of dimensions
\beq
	 \int_{-T/2}^{T/2} f^2(t) dt \leq P N_0, 
\label{eq:ecnS}
\eeq
and adds a zero mean Gaussian noise variable of standard deviation $\sigma$ independently to each coordinate of the space. In this model, the  signal to noise ratio on each coordinate is  
\beq
\SNR_S = P /\sigma^2.
\eeq

Shannon's  capacity is the logarithm base two of the largest number of messages $M_{\sigma}^{\delta}(P)$ that can be communicated  with probability of error $\delta>0$. When taken per unit time,  this is
\beq
C = \lim_{T \rightarrow \infty} \frac{\log M_{\sigma}^{\delta}(P)}{T} \;\;\; \mbox{bits per second},
\label{sdef}
\eeq
and it does not depend on  $\delta$.
The definition in (\ref{sdef}) should be compared with (\ref{cdef}). The geometric insight on which the two models are built upon is the same. However, while in Kolmogorov's deterministic model packing is performed with ``hard'' spheres of radius $\epsilon$ and communication in the presence of arbitrarily distributed noise over a bounded support is performed without error, in Shannon's stochastic model packing is performed with ``soft'' spheres of effective radius $\sqrt{N_0} \sigma$ and communication in the presence of Gaussian noise of unbounded support is performed with arbitrarily low probability of error $\delta$. 

Shannon's energy constraint (\ref{eq:ecnS}) scales with the number of dimensions, rather than being a constant. The reason for this should be clear: since the noise is assumed to act independently on each signal's coefficient,  the statistical spread of the output, given the input signal, corresponds to an uncertainty ball  of radius  $\sqrt{N_0} \sigma$. It follows that the norm of the signal should also be proportional to $\sqrt{N_0}$, to avoid a vanishing signal to noise ratio as $N_0 \rightarrow \infty$.  In contrast, in the case of Kolmogorov the capacity is computed assuming an uncertainty ball of fixed radius $\epsilon$ and the energy constraint is constant. 
In both cases, spectral concentration ensures that the size of the signals' space is essentially of $N_0$ dimensions. Probabilistic concentration ensures that the noise in Shannon's model concentrates around its  standard deviation, so that the functional form of the results is similar in the two cases. 

Shannon's celebrated formula for the capacity of the Gaussian model is~\cite{shannon1948}
\beq
C = \frac{\Omega}{\pi} \log (\sqrt{1+\SNR_S}) \;\;\; \mbox{bits per second}.
\eeq

Our results  for Kolmogorov's deterministic model are  
\begin{empheq}[left=\empheqlbrace]{align}
\bar{C}_{2 \epsilon} &\leq \frac{\Omega}{\pi} \log \left(1+\sqrt{\SNR_K/2}\right) \;\;\; \mbox{bits per second},   \label{uno}\\ 
\bar{C}_{2 \epsilon} & \geq \frac{\Omega}{\pi} \left(\log \sqrt{\SNR_K} -1\right) \label{due} \;\;\; \mbox{bits per second}.
\end{empheq}

The upper bound (\ref{uno}) is an improved version of our previous one in~\cite{allerton1}. For high values of the signal to noise ratio, it   becomes approximately $\Omega/\pi \left( \log \sqrt{\SNR_K}-1/2 \right)$, i.e.\ tight up to a term $\Omega/(2 \pi)$. Both upper and lower bounds are improvements over the ones given by Jagerman~\cite{jagerman1969,jagerman1970}, see Appendix~\ref{a:comparison} for a discussion.

To provide a more precise comparison between the deterministic and the stochastic model, we extend the deterministic model allowing signals in the codebook to be at distance less than $2 \epsilon$ of each other. We say that signals in a codebook are $(\epsilon,\delta)$-distinguishable if the portion of space where the received signal may fall and result in a decoding error is of measure at most $\delta$.  The $(\epsilon,\delta)$-capacity is the  logarithm base two of the maximum number $M_\epsilon^\delta(E)$ of $(\epsilon,\delta)$-distinguishable signals in the space and it is given by
\beq
	C_{\epsilon}^{\delta}   =  \log M_{\epsilon}^{\delta} (E)  \;\;\; 
	\mbox{bits}.
\eeq
We also define the $(\epsilon,\delta)$-capacity per unit time
\beq
\label{cdef'}
	\bar{C}_{\epsilon}^{\delta}   =  
	\lim_{T \rightarrow \infty} \frac{\log M_{\epsilon}^{\delta} (E)}{T} \;\;\; 
	\mbox{bits per second}.\
\eeq
In this case, we show, for any $\epsilon, \delta>0$ 
\begin{empheq}[left=\empheqlbrace]{align}
\bar{C}_{\epsilon}^{\delta} &\leq \frac{\Omega}{\pi} \log \left(1+\sqrt{\SNR_K}\right) \;\;\; \mbox{bits per second},   \label{tre}\\ 
\bar{C}_{\epsilon}^{\delta} & \geq \frac{\Omega}{\pi} \log \sqrt{\SNR_K}  \label{quattro} \;\;\; \mbox{bits per second}.
\end{empheq}

As in Shannon's case, these results do not depend on the size of the error region  $\delta$.  They become tight for high values of the signal to noise ratio.

The lower bound follows from a random coding argument by reducing the problem to the existence of a coding scheme for a stochastic uniform noise channel  with arbitrarily small probability of error. The existence of such a scheme in the stochastic setting implies the existence of a corresponding scheme in the deterministic setting as well. 
Comparing (\ref{uno}) and (\ref{quattro}) it follows that in the high $\SNR_K$ regime, where
\beq
\sqrt{E}>\frac{\sqrt{2}}{\sqrt{2}-1} \epsilon,
\eeq
having a positive error region guarantees a strictly larger capacity. Given our proof reduction, this corresponds to having a Shannon capacity for the uniform noise channel strictly greater than the corresponding zero-error capacity.
  
 The analogy between the size of the error region in the deterministic setting and the probability of error in the stochastic setting also leads to a notion of deterministic error exponent. Letting the number of messages in the codebook be $M=2^{TR}$, where the transmission rate $R$ is smaller than the lower bound (\ref{quattro}), in Appendix~\ref{a:err} we  bound the size of the error region to be at most
 \beq
 \delta \leq 2^{-T\left( \frac{\Omega}{\pi} \log \scriptsize{\sqrt{\SNR_K}} -R \right)},
 \eeq
 and the error exponent in the deterministic model is
 \beq
 \Er(R) =\frac{\Omega}{\pi}\log \sqrt{\SNR_K} -R >0, 
 \eeq
 that depends only on $\Omega$, $E$, $\epsilon$, and on the transmission rate $R$.

\subsection{Entropy}
We consider the same signal space as above, corresponding to points  of essentially $N_0 = \Omega T/\pi$ dimensions and contained in a ball of radius $\sqrt{E}$. A source codebook is composed by a subset of points in this space, and each codebook point is  a possible representation for the signals that are within  radius $\epsilon$ of itself. If the union of the $\epsilon$ balls centered at all codebook points covers the whole space, then any signal  in the space can be encoded by its closest representation. The radius $\epsilon$ of the covering balls provides a bound on the largest estimation error between any source $f(t)$ and its codebook representation $\hat{f}(t)$. When signals are observed over a finite time interval $[-T/2,T/2]$, this corresponds to
\beq
d[f(t),\hat{f}(t)] = \int_{-T/2}^{T/2} [f(t)-\hat{f}(t)]^2 dt \leq \epsilon^2.
\eeq
Following the usual convention in the literature, we call this distortion measure noise, so that the signal to distortion ratio in this source coding model is again  $\SNR_K=\sqrt{E}/\epsilon$.

%


The Kolmogorov $\epsilon$-entropy is the logarithm base two of the minimum number  $L_{\epsilon}(E)$ of $\epsilon$-balls covering the whole space and it is given by
\beq
	H_{\epsilon}   =  \log L_{\epsilon} (E) \;\;\; 
	\mbox{bits}.
\eeq
We also define the $\epsilon$-entropy per unit time
\beq
\label{hdef}
	\bar{H}_{\epsilon}   =  
	\lim_{T \rightarrow \infty} \frac{\log L_{\epsilon} (E)}{T} \;\;\; 
	\mbox{bits per second}.\
\eeq

An analogous Gaussian stochastic source model, due to Shannon, models the source signal as  a white Gaussian stochastic process of constant power spectral density $P$  of support $[-\Omega, \Omega]$. This stochastic process has infinite energy, and finite average power 
\beq
\mathds{E}(\f^2(t))=R_\f(0) = \frac{1}{2\pi} \int_{-\infty}^{\infty} S_\f(\omega) d \omega =  \frac{P \Omega}{\pi},
\eeq
where $R_\f$ and $S_\f$ are the autocorrelation and the power spectral density of  $f(t)$, respectively. When observed  over the interval $[-T/2,T/2]$, the process can be 
viewed as a random point having essentially $N_0$ independent Gaussian coordinates of zero mean and variance $P$, and of energy
\beq
\int_{-T/2}^{T/2} \mathds{E}(\f^2(t)) dt = \frac{P \Omega T}{\pi} = P N_0.
\eeq
A source codebook is composed by a subset of points in the space, and each codebook point is  a possible representation for the stochastic process. The distortion associated to the representation of   $f(t)$ using codebook point $\hat{f}(t)$ is defined in terms of mean-squared error 
\beq
d[\f(t),\hat{\f}(t)] = \int_{-T/2}^{T/2}\mathds{E} [\f(t)-\hat{\f}(t)]^2 dt.
\eeq
Letting $L_{\sigma}(P)$  be the smallest number of codebook points that can be used to represent the source process  with distortion at most $\sigma^2 N_0$,  the rate-distortion function is defined as
\beq
	R_{\sigma} = \lim_{T \rightarrow \infty} \frac{\log L_{\sigma}(P)}{T} \;\;\; \mbox{bits per second}.
\label{rdef}
\eeq

In this setting, Shannon's formula for the rate distortion function of a Gaussian source is~\cite{shannon1948}
\beq
	R_{\sigma} = \frac{\Omega}{\pi} \log (\sqrt{\SNR_S}) \;\;\; \mbox{bits per second}.
\eeq

We show the corresponding result in Kolmogorov's deterministic setting
\beq
	\bar{H}_{\epsilon} = \frac{\Omega}{\pi} \log (\sqrt{\SNR_K}) \;\;\; \mbox{bits per second}.
	\label{eqentropyresult}
\eeq
Previously, Jagerman~\cite{jagerman1969, jagerman1970} has shown
\beq
0 \leq \bar{H}_{\epsilon} \leq \frac{\Omega}{\pi} \log \left(1+2\sqrt{\SNR_K}\right), 
\eeq
see Appendix~\ref{a:comparison} for a discussion.
Our result in (\ref{eqentropyresult}) can be derived by combining  a theorem of  Dumer, Pinsker and Prelov~\cite[Theorem~2]{Dumer2004}, on the thinnest covering of ellipsoids in Euclidean spaces of arbitrary dimension, our  Lemma \ref{lemma1}, on the phase transition of the dimensionality of bandlimited square-integrable functions, and an approximation argument given in our Theorem~\ref{thm6}. Instead,  we provide a self-contained proof.


\subsection{Summary}
Table~\ref{table} 
\begin{table*}
\caption{Comparison of stochastic and deterministic models}
\label{table}
\centering  
	\begin{tabular}{l|ll}
	\hline \hline
	&Stochastic   & Deterministic \\
	\hline  \\
	Transmitted Signal & $\int_{-T/2}^{T/2} f^2(t) dt \leq P N_0$ & $\int_{-\infty}^{\infty} f^2(t) dt \leq E$	 \\[.1cm]
	Additive Noise 	  & $\mathds{E} \sum_{i=1}^{N_0}  \mathsf{n}_i^2 =  N_0 {\sigma}^2$ & $ \sum_{i=1}^{\infty} n_i^2 \leq {\epsilon}^2$	 \\[.1cm]
	 Effective Dimensionality & $N_0 = \Omega T/\pi$ & $N_0 = \Omega T/\pi$		\\[.1cm]		
	Signal to Noise Ratio  &$ \mathsf{SNR}_S = P / \sigma^2$ & $\mathsf{SNR}_K= E/\epsilon^2$ \\[.1cm]
	Max Cardinality of Codebook   & $M_{\sigma}^{\delta}(P)$& $M_{\epsilon}^{\delta}(E)$ \\[.1cm]  
	Capacity & $C = \frac{\Omega}{\pi} \log (\sqrt{1+ \SNR_S})$ 	&	$\frac{\Omega}{\pi} \log \sqrt{\SNR_K} \leq {\bar{C}}_{\epsilon}^{\delta} \leq \frac{\Omega}{\pi} \log (1+\sqrt{\SNR_K})$
	\\ \\ \hline   \\ 
	Source Signal & $\int_{-T/2}^{T/2} \mathds{E}(\f^2(t)) dt = P N_0$ 							& $\int_{-\infty}^{\infty} f^2(t) dt \leq E$	\\[.1cm]
	Distortion & $d[\f(t),\hat{f}(t)]  \leq N_0 \sigma^2$ & $d[\f(t),\hat{\f}(t)]  \leq  \epsilon^2$\\[.1cm]
	Min Cardinality of Codebook  & $L_{\sigma}(P)$& $L_{\epsilon}(E)$\\[.1cm]
	Rate Distortion Function & $R_{\sigma}= \frac{\Omega}{\pi} \log \sqrt{\SNR_S}$ 						& ${\bar{H}}_{\epsilon} = \frac{\Omega}{\pi} \log \sqrt{\SNR_K}$
	\\ \\
		\hline	\hline
	\end{tabular}
\end{table*}
provides a comparison between  results in the deterministic and  in the stochastic setting. 
In the computation of capacity, a transmitted signal subject to a given energy constraint, is corrupted by additive noise. Due to spectral concentration, the signal has an effective number of dimensions $N_0$. In a deterministic setting, the noise  represented by   the deterministic coordinates $\{n_i\}$, can take any value inside a ball of radius $\epsilon$.  In a stochastic setting, due to probabilistic concentration, the noise  represented by the stochastic coordinates $\{\mathsf{n}_i\}$, can take   values essentially uniformly at random inside a ball of effective radius $N_0 \sigma^2$.    In both cases, the maximum cardinality of the codebook used for communication depends on the error measure $\delta>0$, but the capacity in bits per unit time does not, and it depends only on the signal to noise ratio. The special case $\delta=0$ is treated separately, and it does not appear in the table. This corresponds to the  Kolmogorov $2 \epsilon$-capacity, and is the analog of the Shannon zero-error capacity of an  $\epsilon$-bounded noise channel.

In the computation of the rate distortion function, a source signal is modeled as either an arbitrary, or stochastic process of given energy constraint. The distortion measure corresponds to the estimation error incurred when this signal is represented by an element of the source codebook. The minimum cardinality of the codebook used for representation depends on the distortion constraint, and so does the rate distortion function. 

In both the deterministic and stochastic settings we  have a tight asymptotic characterization of the rate distortion function, while we  have bounds for the capacity in the deterministic setting that are tight only in the high $\textsf{SNR}_K$ regime. This is because distances in the probabilistic model are measured in terms of standard deviation, while they are measured in terms of $L^2[-T/2,T/2]$ norm in the deterministic model. The computation of capacity requires to sum the signal and the noise, and in the probabilistic model  the norm of the sum of two signals can be expressed as the square root of the sum of their variances, leading to a tight expression. In the deterministic model, the norm of the sum of two signals can only be bounded, and  this leads to a gap between upper and lower bounds that vanishes for high values of $\textsf{SNR}_K$. In the case of rate distortion,  we do not need to compute the sum of two signals, and tight bounds are obtained in both settings.


\section{The signals' space}
\label{sec:problem}

We now describe  the signals' space rigorously,  mention some classic results required for our derivations,  introduce rigorous notions of capacity and entropy, and present the technical approach that we use  in the proofs.

\subsection{Energy-constrained, bandlimited functions}

We consider the  set of one-dimensional, real, bandlimited functions  
\beq
\label{eq:def1-1}
	\BO = \{ f(t) : \mathcal{F}f(\omega)  = 0, \mbox{ for } |\omega| > \Omega \},
\eeq
where
\beq
 \mathcal{F}f(\omega)=  \int_{-\infty}^{\infty} f(t) \exp(-j \omega t) dt,
\eeq
and $j$ denotes the imaginary unit.

These functions are assumed to be square-integrable, and to satisfy the energy constraint (\ref{eq:ecn}).
We equip them with the $L^2[-T/2,T/2]$ norm
\beq
\label{eq:normdef1}
	\|f\|= \left( \int_{-\frac{T}{2}}^{\frac{T}{2}} f^2(t) dt \right)^{1/2}.
\eeq
It follows that $(\BO, \|\cdot\|)$ is a metric space, whose elemets are 
real, bandlimited functions, of infinite duration and observed over a finite interval $[-T/2,T/2]$. 
The elements of this space can be optimally approximated, in the sense of Kolmogorov, using a finite series expansion of a suitable basis set.

\subsection{Prolate spheroidal basis set}

Given any $T, \Omega>0$, there exists a countably infinite set of real functions
$\{ \psi_n(t) \} _{n=1} ^{\infty}$, called prolate spheroidal wave functions (PSWF), 
and a set of real positive numbers $1>\lambda_1>\lambda_2>\cdots$ with the following properties:

{\it{Property 1.}}
The elements of $\{\lambda_n\}$ and $\{\psi_n\}$ are solutions of the  Fredholm integral  equation of the second kind
\beq
	\lambda_n \psi_n (t) = \int_{-\frac{T}{2}}^{\frac{T}{2}} \psi_n(s) \frac{\sin \Omega (t-s)}{\pi(t-s)} ds.
	\label{eq:fredholm}
\eeq

{\it{Property 2.}}
The elements of $\{\psi_n(t)\}$ have Fourier transform that is zero for $|\omega|>\Omega$.

{\it{Property 3.}}
The set $\{ \psi_n(t) \} $ is complete in $\mathcal{B}_{\Omega}$. 

{\it{Property 4.}}
The elements of $\{ \psi_n(t) \}$  are orthonormal in   $(-\infty, \infty)$ 
\beq
\label{p4}
	\int_{-\infty}^{\infty} \psi_n(t) \psi_m(t) dt = 
		\begin{cases}
		1 & n=m,\\
		0 & \mbox{otherwise}.
		\end{cases}
\eeq

{\it{Property 5.}}
The elements of $\{ \psi_n(t) \}$ are orthogonal in  $\left(-\frac{T}{2}, \frac{T}{2} \right)$ 
\beq
\label{p5}
	\int_{-\frac{T}{2}}^{\frac{T}{2}} \psi_n(t) \psi_m(t) dt = 
		\begin{cases}
		\lambda_n & n=m,\\
		0 & \mbox{otherwise}.
		\end{cases}
\eeq

{\it{Property 6.}}
The eigenvalues in $\{\lambda_n\}$ undergo a phase transition at the scale of $N_0 = \Omega T/\pi$: 
for any $\alpha>0$ 
\beq
{\label{p_6-1}}
	\lim_{N_0 \rightarrow \infty} \lambda_{\lfloor (1-\alpha) N_0 \rfloor} =1,
\eeq
\beq
{\label{p_6-2}}
	 \lim_{N_0\rightarrow \infty} \lambda_{\lfloor (1+\alpha)N_0 \rfloor}=0.
\eeq

{\it{Property 7.}}
The width of the phase transition can be precisely characterized:
for any $k>0$ 
\beq
{\label{p_7}}
 \lim_{N_0\rightarrow \infty} \lambda_{\lfloor N_0 + k \log (N_0 \pi/2) \rfloor}  = \frac{1}{1+e^{k\pi^2}}.
\eeq

For an extended treatment of PSWF see \cite{flammer2005}. 
The phase transition behavior of the eigenvalues is a key property related to  the number of terms required   for a satisfactory approximation of any  square integrable bandlimited function using a finite basis set. Much of the theory was developed jointly by Landau, Pollack, and Slepian, see \cite{slepian1983} for a review. The precise asymptotic behavior in (\ref{p_7}) was finally proven by Landau and Widom~\cite{landau1980}, after a conjecture of Slepian supported by a non-rigorous computation~\cite{slepian1965}.

\subsection{Approximation of $\BO$}
Let  $\mathcal{X}=L^2[-T/2,T/2]$, 
the Kolmogorov $N$-width~\cite{pinkus} of $\BO$ in $\mathcal{X}$ is  
\begin{equation}
	d_N(\BO,\mathcal{X})=\inf_{\mathcal{X}_N \subseteq \mathcal{X}} \sup_{f \in \BO} \inf_{g \in \mathcal{X}_N}
	\|f-g\|,
\end{equation}
where $\mathcal{X}_N$ is an   $N$-dimensional suspace of $\mathcal{X}$. For any $\mu>0$,
we use this notion to define the number of degree of freedom at level $\mu$ of the space $\BO$  as 
\begin{equation}
	N_{\mu}(\BO) = \min\{N : d_{N}(\BO, \mathcal{X}) \leq \mu \}.
	\label{ndf}
\end{equation}

In words, the Kolmogorov $N$-width represents the extent to which $\BO$ may be uniformly approximated by  an $N$-dimensional  subspace of $\mathcal{X}$, and the number of degrees of freedom is the dimension of the minimal subspace representing the elements of $\BO$ within the desired accuracy $\mu$. It follows that the number of degrees of freedom represents the effective dimensionality of the space, and corresponds to the number of coordinates that are essentially needed to identify any one element in the space.

A basic result in approximation theory (see e.g.\ \cite[Ch. 2,  Prop. 2.8]{pinkus}) states that
\begin{equation}{\label{p1}}
	d_N (\BO, \mathcal{X}) = \sqrt{E \lambda_{N+1}},
\end{equation}
and the corresponding approximating subspace is the one spanned by the PSWF basis set $\{\psi_n\}_{n=1}^{N}$. It follows that any
 bandlimited function $f \in \BO$ can be optimally approximated by retaining a finite number $N$ of terms in the series expansion 
\beq
	f(t) = \sum_{n=1}^{\infty} b_n \psi_n (t),
	\label{proexp}
\eeq
and  that the number of degree of freedom in (\ref{ndf}) is    given by the minimum index $N$  such that $\sqrt{\lambda_{N+1}} \leq \mu/\sqrt{E}$.
The phase transition of the eigenvalues  ensures that this  number is only slightly larger than $N_0$.
More precisely,
for any  $\mu>0$  we may  choose an integer 
\begin{align}
N 
&= N_0 +  \frac{1}{\pi^2} \log \left( \frac{E}{\mu^2} -1\right) \log \left(\frac{N_0 \pi}{2} \right) +o(\log N_0),
\label{eq:choose}
\end{align}
and approximate    
\beq
\BO= \left\{ {\bf{b}} = (b_1, b_2, \cdots ) : \sum_{n=1}^{\infty}b_n^2 \leq E \right\},
	\label{bo}
\eeq
 within accuracy $\mu$ as $N_0 \rightarrow \infty$ using  
\beq
	\BOP = \left\{ {\bf{b}} = (b_1, b_2, \cdots, b_N) : \sum_{n=1}^{N}b_n^2 \leq E \right\},
	\label{bop}
\eeq
equipped with the norm
\beq
	\|{\bf{b}}\|' = \sqrt{\sum_{n=1}^{N} b_n^2 \lambda_n}.
	\label{eq:metric1}
\eeq

The energy constraint in (\ref{bop}) follows from (\ref{eq:ecn}) using the orthonormality Property 4 of the PSWF,  the norm in (\ref{eq:metric1}) follows from (\ref{eq:normdef1}) using the orthogonality Property 5 of the PSWF, and the desired level of approximation is guaranteed by Property 7 of the PSWF. 

By  (\ref{eq:choose}) it follows that the number of degrees of freedom is an  intrinsic property of the space, essentially dependent on the time-bandwidth product $N_0 = \Omega T/\pi$, and only weakly, i.e. logarithmically,  on the accuracy $\mu$ of the approximation and on the energy constraint $E$.

These approximation-theoretic results show that any energy-constrained, bandlimited waveform can be identified by essentially $N_0$ real numbers. This  does not pose a limit on the amount of information carried by the signal. The real numbers identifying the waveform can be specified up to arbitrary precision, and this results in an infinite number of possible waveforms that can be used for communication. 
To bound the amount of information, we need to introduce a resolution limit at which the waveform can be observed, which allows  an information-theoretic description of the space using bits rather than  real numbers. This description is given in terms of entropy and capacity.
\subsection{$\epsilon$-entropy and $\epsilon$-capacity}

\begin{figure}[!t]
\centering
\includegraphics[width=80mm]{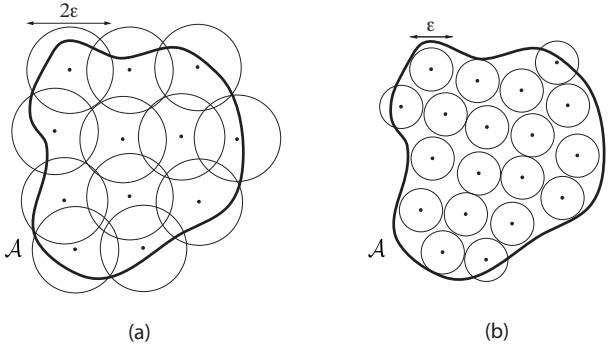}
\caption{Part (a): Illustration of the $\epsilon$-entropy. Part (b): Illustration of the $\epsilon$-capacity.}
\label{fig1}
\end{figure}

Let $\mathcal{A}$ be a subset of the metric space $\mathcal{X}=L^2[-T/2,T/2]$. A set of points  in $\mathcal{A}$ is called an $\epsilon$-covering  if for any point   in $\mathcal{A}$ there exists a point in the covering at distance at most $\epsilon$ from it. The minimum cardinality of an $\epsilon$-covering  is an invariant of the set $\mathcal{A}$, which depends only on $\epsilon$, and is denoted by $L_\epsilon(\mathcal{A})$. The $\epsilon$\emph{-entropy} of $\mathcal{A}$ is defined as the base two logarithm
\beq
	H_{\epsilon}(\mathcal{A}) = \log L_\epsilon(\mathcal{A}) \;\;\; 
	\mbox{bits},\
\eeq
see Fig. {\ref{fig1}}-(a). 
We also define the $\epsilon$-entropy per unit time
\beq
	{\bar{H}}_{\epsilon}(\mathcal{A}) = 
	\lim_{T \rightarrow \infty} \frac{H_{\epsilon}(\mathcal{A})}{T} \;\;\; 
	\mbox{bits per second}.\
\eeq

A set of points   in $\mathcal{A}$  is called $\epsilon$-distinguishable if the distance between any two of them exceeds $\epsilon$. The maximum cardinality of an $\epsilon$-distinguishable set  is an invariant of the set $\mathcal{A}$, which depends only on $\epsilon$, and is denoted by $M_\epsilon(\mathcal{A})$. The $\epsilon$\emph{-capacity} of $\mathcal{A}$ is defined as the base two logarithm
\beq
	C_\epsilon(\mathcal{A}) = \log M_\epsilon(\mathcal{A})  \;\;\; 
	\mbox{bits},\
\eeq
see Fig. {\ref{fig1}}-(b).
We also define the $\epsilon$-capacity per unit time
\beq
	{\bar{C}}_{\epsilon}(\mathcal{A}) = 
	\lim_{T \rightarrow \infty} \frac{C_{\epsilon}(\mathcal{A})}{T} \;\;\; 
	\mbox{bits per second}.\
\eeq

\begin{figure}[!t]
\centering
\includegraphics[width=80mm]{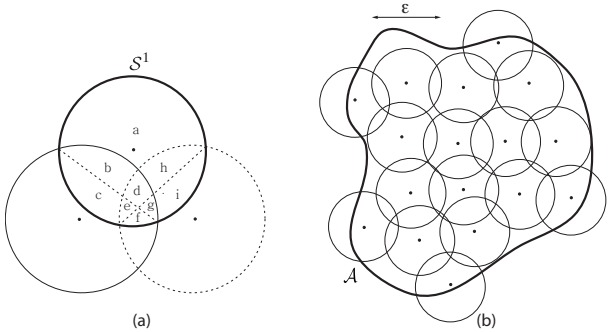}
\caption{Part (a): Illustration of the error region for a signal in the space.  The letters indicate the volume of the corresponding regions of the ball $\mathcal{S}^1$, and   $\Delta_1 = (c+e+f+g+i)/(a+b+c+d+e+f+g+h+i)$.  Part (b): Illustration of the $(\epsilon,\delta)$-capacity. An overlap among the $\epsilon$-balls is allowed, provided that the cumulative error measure $\Delta \leq \delta$.}
\label{fig2}
\end{figure}

The $\epsilon$-entropy and $\epsilon$-capacity are  closely related to the probabilistic notions
of entropy and capacity used in information theory. 
The $\epsilon$-entropy corresponds  to the rate distortion function, and the $\epsilon$-capacity corresponds to the zero-error capacity. 
In order to have a deterministic quantity that corresponds to the Shannon capacity, we extend the $\epsilon$-capacity and allow  a small fraction of  intersection among the $\epsilon$-balls when constructing a packing set. This leads to a certain region of space where the received signal may fall and result in a communication error, and to the notion of $(\epsilon, \delta)$-capacity.  

\subsection{$(\epsilon,\delta)$-capacity}


Let $\mathcal{A}$ be a subset of the metric space  $\mathcal{X}=L^2[-T/2,T/2]$. We consider a set of points in $\mathcal{A}$, $\mathcal{M}=\{ {\bf{a}}^{(1)}, {\bf{a}}^{(2)}, \cdots, {\bf{a}}^{(M)} \}$. For a given ${\bf{a}}^{(i)}$, $1\leq i \leq M$, we let the noise ball
\beq
	\mathcal{S}^{i} =  \{ {\bf{x}} \in \mathcal{X} : \| {\bf{x}}-{\bf{a}}^{(i)} \| \leq \epsilon \},
\eeq
where $\epsilon$ is a positive real number, and we let error region with respect to minimum distance decoding
\beq
	\mathcal{D}^{i} =  \{ {\bf{x}} \in \mathcal{S}^{i} : \exists j\not =i :  
	\| {\bf{x}}-{\bf{a}}^{(j)} \| \leq \| {\bf{x}}-{\bf{a}}^{(i)} \|\}.
\eeq
We define the error measure for the $i$th signal
\beq
	\Delta_i = \frac{{\rm{vol}}(\mathcal{D}^i)}{{\rm{vol}}(\mathcal{S}^i)},
\label{di}
\eeq
where ${\rm{vol}(\cdot)}$ indicates volume in $\mathcal{X}$, and the cumulative error measure
\beq
	\Delta = \frac{1}{M} \sum_{i=1}^{M}{\Delta_i},
\label{cumdi}
\eeq
Fig. {\ref{fig2}}-(a) provides an illustration of the error region for a signal in the space.
Clearly, we have $0 \leq \Delta \leq 1$. For any $\delta>0$, we say that a set of points $\mathcal{M}$ in $\mathcal{A}$ is $(\epsilon,\delta)$-distinguishable set if $\Delta \leq \delta$.  The maximum cardinality of an $(\epsilon,\delta)$-distinguishable set is an invariant of the space $\mathcal{A}$, which depends only on $\epsilon$ and $\delta$, and is denoted by $M_{\epsilon}^{\delta}(\mathcal{A})$. The $(\epsilon,\delta)$-capacity of $\mathcal{A}$ is defined as the base two logarithm
\beq
	C_{\epsilon}^{\delta} (\mathcal{A}) =  \log M_{\epsilon}^{\delta} (\mathcal{A})  \;\;\; 
	\mbox{bits},
\eeq
see Fig. {\ref{fig2}}-(b).
We also define the $(\epsilon,\delta)$-capacity per unit time
\beq
	\bar{C}_{\epsilon}^{\delta} (\mathcal{A}) =  
	\lim_{T \rightarrow \infty} \frac{C_{\epsilon}^{\delta}({\mathcal{A}})}{T} \;\;\; 
	\mbox{bits per second}.\
\eeq

\subsection{Technical approach}
Our objective is to compute entropy and capacity of square integrable, bandlimited functions. 
First, we perform this computation for the finite-dimensional space of functions $\BOP$ that approximates   the infinite-dimensional space $\BO$   up to arbitrary accuracy $\mu>0$ in the $L^2[-T/2,T/2]$ norm, as $N_0 \rightarrow \infty$. 
Our results in this setting are given by Theorem~\ref{thm1} for the $\epsilon$-capacity,  Theorem~\ref{thm2} for the $(\epsilon,\delta)$-capacity, and Theorem~\ref{thm3} for the $\epsilon$-entropy. Then,  in Theorems~\ref{thm4}, \ref{thm5}, and~\ref{thm6}, we extend the computation to  the $\epsilon$-capacity, $(\epsilon,\delta)$-capacity, and $\epsilon$-entropy    of the whole space $\BO$ of bandlimited functions. 

When viewed per unit time, results for the two spaces are identical, indicating   that using a highly accurate, lower-dimensional subspace approximation   leaves only a negligible  ``information leak'' in higher dimensions.  We bound this leak  in the case of $\epsilon$-entropy and $\epsilon$-capacity by performing a projection from the high-dimensional space $\BO$ onto the lower-dimensional one $\BOP$ and noticing that distances do not change significantly when these two spaces are sufficiently close to one another. 
On the other hand,  for the $(\epsilon, \delta)$-capacity   the error is defined in terms of volume, which may change significantly, no matter how close the two spaces are. In this case, we cannot bound the $(\epsilon, \delta)$ capacity of  $\BO$   by performing a projection onto $\BOP$, and instead provide a  bound  on the capacity per unit time in terms  of another finite-dimensional space
that asymptotically approximates $\BO$ with   perfect accuracy $\mu=0$, as $N_0 \rightarrow \infty$. 
\section{Applications}
Recent interest in deterministic models of information has been raised in the context of control theory and electromagnetic wave theory. 

Control theory often treats uncertainties and disturbances as bounded unknowns having no statistical structure.  In this context,  Nair~\cite{nair2013}  introduced a maximin information functional for non-stochastic variables and  used it to derive tight conditions for uniformly estimating the state of a linear time-invariant system over an error-prone channel. The relevance of Nair's approach  to estimation over unreliable channels is due to its connection with the Shannon zero-error capacity~\cite[Theorem 4.1]{nair2013}, which  has applications in networked control theory~\cite{matveev}.  In Appendix~\ref{a:nair} we point out that Nairs' maximum information rate functional, when viewed in our continuous setting of communication with bandlimited signals, is nothing else than $\bar{C}(\BO)$. This suggests that our approach can be used in the same context as his. 

In electromagnetic wave theory,  the field measurement accuracy, and the corresponding image resolution in remote sensing applications, are often treated as fixed constants below which distinct electromagnetic fields, corresponding to different images, must be considered indistinguishable. In this framework, the number of degrees of freedom of radiating fields has been determined starting from their bandlimitation properties~\cite{bucci1, bucci2}.  Using the same approach, in communication theory the number of parallel channels available  in spatially distributed multiple antenna systems under a fixed noise level constraint has been  determined and related to the cut-set boundary separating transmitters and receivers~\cite{massimo2014}.
Our results can be used in the same setting to provide the extension from the approximation-theoretic notion of degrees of freedom to the information-theoretic ones of entropy and capacity, something already suggested in~\cite{bucci2}.

Several other   applications of the deterministic approach pursued here seem worth exploring, including the analysis of multi-band signals of sparse support. More generally, one could study capacity and entropy under different  constrains beside bandlimitation, and attempt, for example, to obtain  formulas analogous to waterfilling solutions in a deterministic setting.
\section{Nothing but proofs}
\label{sec:main}


We start with some preliminary lemmas that are needed for the proof of our main theorems. The first lemma is a consequence of  the phase transition of the eigenvalues, while the second and third lemmas are properties of Euclidean spaces.


\begin{lemma}
Let
\beq
	\zeta(N)=\left( \prod_{i=1}^{N} \lambda_{i} \right)^{1/(2N)},
\eeq
where $N=N_0 +O(\log N_0)$ as $N_0 \rightarrow \infty$. We have
\beq
	\lim_{N_0 \rightarrow \infty} \zeta(N) = 1.
\eeq
\label{lemma1}
\end{lemma}
\begin{IEEEproof}
For any $\alpha>0$, we  have
\begin{align}
	\log \zeta(N) 
	&= \frac{1}{2N} \sum_{i=1}^{N} \log \lambda_{i} \nonumber \\
	&= \frac{1}{2N} \left( \sum_{i=1}^{\lfloor (1-\alpha)N_0 \rfloor} \log \lambda_{i}  \right. \nonumber \\
	& \left. \hspace{.7cm} + \sum_{i=\lfloor (1-\alpha)N_0 \rfloor+1}^{N} \log \lambda_{i} \right).
	\label{eq:secondsum}
\end{align}
From  Property 6 of the PSWF and the monotonicity of the eigenvalues it follows that the first sum in (\ref{eq:secondsum}) tends to zero as $N_0 \rightarrow \infty$. We turn our attention to the second sum.  
By the monotonicity of the eigenvalues, we have
\beq
	\sum_{i=\lfloor (1-\alpha)N_0+1\rfloor}^{N} \log \lambda_{i}  \geq (N-(1-\alpha)N_0) \log \lambda_{N}.
\eeq

Since $N = N_0 + O(\log N_0)$ as $N_0 \rightarrow \infty$, there exists a constant $k$ such that for $N_0$ large enough  $N \leq N_0 + k \log N_0$ and the right-hand side is an integer. 
It follows that for $N_0$ large enough, we have
\begin{align}
	\sum_{i=\lfloor (1-\alpha)N_0+1\rfloor}^{N} \log \lambda_{i} 
	&\geq (\alpha N_0 + k \log N_0) \log \lambda_{N} \nonumber \\
	& \geq (\alpha N_0 + k \log N_0)  \nonumber \\&\times \log(\lambda_{N_0 + k \log N_0}).
	\label{eq:lowerb}
\end{align}

Substituting (\ref{eq:lowerb}) into (\ref{eq:secondsum}) and using Property 7 of the PSWF, it follows that for $N_0$ large enough

\begin{align}
	\log \zeta(N) &\geq \frac{\alpha N_0 + k \log N_0}{2N} \log \left( \frac{1}{1+e^{\pi^2 k}} \right),
\end{align}
and since $N = N_0 + O(\log N_0)$ as $N_0 \rightarrow \infty$, we have
\beq
	\lim_{N_0 \rightarrow \infty} \log \zeta(N) \geq \frac{\alpha}{2}\log \left( \frac{1}{1+e^{\pi^2 k}} \right).
\eeq
The proof is completed by noting that $\alpha$ can be arbitrarily small.
\end{IEEEproof}

\begin{lemma}
\label{lemma2}
Let $m$ be a positive integer and let ${\bf{x}}, {\bf{x}}^{(1)}, \cdots, {\bf{x}}^{(m)}$ be arbitrary points in $n$-dimensional Euclidean space, $(\mathbb{E}^n$, $\| \cdot \|)$. We have
\beq
	\sum_{j=1}^{m}\sum_{k=1}^{m} \| {\bf{x}}^{(j)} - {\bf{x}}^{(k)} \|^2
	\leq
	2m \sum_{j=1}^{m} \| {\bf{x}} - {\bf{x}}^{(j)} \|^2.
\eeq
\end{lemma}
The proof is given  in \cite[Lemma 6.1]{Zong2013}.

\begin{lemma}
\label{lemma3}
Let $L$ be the cardinality of the minimal $\epsilon$-covering of the $\sqrt{E}$-ball   $\mathcal{S}_{\sqrt{E}}$  in $\mathbb{E}^n$.  If $n\geq 9$, we have
\beq
	L \leq \frac{4e \cdot n^{3/2} \left( \frac{\sqrt{E}}{\epsilon} \right)^n }{\ln{n}-2} \left[ n \cdot \ln{n} + o(n \cdot \ln{n}) \right]
\eeq
where $1< \frac{\sqrt{E}}{\epsilon} < \frac{n}{\ln{n}}$.
\end{lemma}
The proof is given in \cite[Theorem 2]{Rogers1963}.

\subsection{Main theorems for $\BOP$}


Although the set $\BOP$ in (\ref{bop}) defines an $N$-dimensional hypersphere, the metric in (\ref{eq:metric1}) is not Euclidean. It is convenient to express   the metric in Euclidean form by performing a scaling transformation of the coordinates of the space.  For all $n$, we let
 $a_n = b_n \sqrt{\lambda_n}$, so that we have
\beq
	\BOP = \left\{ {\bf{a}} = (a_1, a_2, \cdots, a_N) : \sum_{n=1}^{N} \frac{a_n^2}{\lambda_n} \leq E \right\}
	\label{bop'}
\eeq
and
\beq
	\|{\bf{a}}\|' = \sqrt{\sum_{n=1}^{N} a_n^2}.
	\label{eq:metric2}
\eeq
We   now consider packing  and covering with $\epsilon$-balls inside the ellipsoid defined in~(\ref{bop'}), using the Euclidean metric in (\ref{eq:metric2}).  



\begin{theorem}
\label{thm1}
	For any $\epsilon>0$, we have
	\begin{empheq}[left=\empheqlbrace]{align}
		 &\bar{C}_{\epsilon}^{0} (\BOP) \geq \frac{\Omega}{\pi} \left[ \log \left(  \frac{\sqrt{E}}{\epsilon} \right) -1 \right], \\
		 &\bar{C}_{\epsilon}^{0} (\BOP) \leq	\frac{\Omega}{\pi} \left[ \log \left( 1+  \frac{\sqrt{E}}{\epsilon \sqrt{2}}  \right) \right].
	\end{empheq}
\end{theorem}

\begin{IEEEproof}
To prove the result it is enough to show the following inequalities for the $2\epsilon$-capacity
\begin{empheq}{align}
	&C_{\epsilon}^{0} ( \BOP) \geq	N \left[ \log \left( \zeta(N) \frac{\sqrt{E}}{\epsilon} \right) -1 \right] \label{eq:lower1}, \\
	&C_{\epsilon}^{0} ( \BOP) \leq N \left[ \log \left( 1+ \frac{\sqrt{E}}{\sqrt{2}\epsilon} \right) \label{eq:upper1}
	\right] + \log \left( 1+\frac{N}{2} \right),
\end{empheq}
because $\lim_{T \rightarrow \infty} \zeta(N) = 1$ and $\log \left( 1+\frac{N}{2} \right) = o(T)$. 


\noindent {\emph{Lower bound.}} \;\;
Let $\mathcal{M}_{\epsilon}^{0}$ be a maximal $(\epsilon,0)$-distinguishable subset of $\BOP$ and $M_{\epsilon}^{0} (\BOP)$ be the number of elements in $\mathcal{M}_{\epsilon}^{0}$. For each point of $\mathcal{M}_{\epsilon}^{0}$, we consider an Euclidean ball whose center is the chosen point and whose radius is $2\epsilon$. Let $\mathcal{U}$ be the union of these balls. We claim that $\BOP$ is contained in $\mathcal{U}$. If that is not the case, we can find a point of $\BOP$ which is not contained in $\mathcal{M}_{\epsilon}^{0}$, but whose distance from every point in $\mathcal{M}_{\epsilon}^{0}$ exceeds $2\epsilon$, which is a contradiction. Thus, we have the  chain of inequalities 
\beq
	{\rm{vol}}(\BOP) \leq {\rm{vol}}(\mathcal{U}) \leq 
	M_{\epsilon}^{0}(\BOP)  {\rm{vol}}(\mathcal{S}_{2\epsilon}),
	\label{eq:S1}
\eeq
where $\mathcal{S}_{2\epsilon}$ is an Euclidean ball whose radius is $2\epsilon$ and the second inequality follows from a union bound. Since ${\rm{vol}}(\mathcal{S}_{\epsilon})= \beta_N \cdot \epsilon^N$, where $\beta_N$ is the volume of $\mathcal{S}_{1}$, by (\ref{eq:S1}) we have
\beq
	{\left( \frac{1}{2} \right)}^{N} \frac{{\rm{vol}}(\BOP)}{{\rm{vol}}(\mathcal{S}_{\epsilon})}
	\leq M_{\epsilon}^{0} (\BOP).
	\label{eq:c1}
\eeq

Since  $\BOP$ is an ellipsoid of radii  $\{\sqrt{\lambda_i E}\}_{i=1}^N$, we also have
\beq
	{\rm{vol}}(\BOP)= \beta_N \prod_{i=1}^N \sqrt{\lambda_i E}  = \beta_N \left( \zeta(N) \sqrt{E} \right)^N,
\eeq
and
\beq
	\frac{  {{\rm{vol}}(\BOP)}  }{  {{\rm{vol}}(\mathcal{S}_{\epsilon})}  } 
	= \left( \zeta(N) \frac{\sqrt{E}}{\epsilon} \right)^{N}.   
	\label{eq:c2}
\eeq

By combining (\ref{eq:c1}) and (\ref{eq:c2}), we get
\beq
	C_{\epsilon}^{0}(\BOP) = \log M_{\epsilon}^{0}(\BOP)\geq	N \left[ \log \left( \zeta(N) \frac{\sqrt{E}}{\epsilon} \right) -1 \right].
\eeq

\noindent
{\emph{Upper bound.}} \;\;
We define the auxiliary set  
\beq
	\BBOP = \left\{ {\bf{a}} = (a_1, a_2, \cdots, a_N) : \sum_{n=1}^{N} a_n^2 \leq E \right\}.
	\label{bop1}
\eeq
The corresponding space $(\BBOP, \|\cdot\|')$ is  Euclidean. Since $\BOP \subset \BBOP$, it follows that $C_{\epsilon}^{0}(\BOP) \leq C_{\epsilon}^{0}(\BBOP)$ and it is sufficient to derive an upper bound for $C_{\epsilon}^{0}(\BBOP)$.

Let $\mathcal{M}_{\epsilon}^{0} = \{ {\bf{a}}^{(1)} , {\bf{a}}^{(2)}, \cdots ,{\bf{a}}^{(M)} \}$ be a maximal $(\epsilon,0)$-distinguishable subset of $\BBOP$, where $M=M_{\epsilon}^{0}(\BBOP)$. 
Let $\{  {\bf{a}}^{(i_1)} , {\bf{a}}^{(i_2)} \cdots  {\bf{a}}^{(i_m)} \}$ be any subset of $\mathcal{M}_{\epsilon}^{0}$.
For any integer $j \neq k$, $j,k \in \{1, \ldots m\}$, we have 
\beq
\| {\bf{a}}^{(i_j)} - {\bf{a}}^{(i_k)}\|' \geq 2\epsilon,
\eeq
and   
\beq
	\sum_{j=1}^{m}\sum_{k=1}^{m} \| {\bf{a}}^{(i_j)} - {\bf{a}}^{(i_k)} \|'^2 
	\geq
	4 {\epsilon}^2 m(m-1).
\eeq
By Lemma \ref{lemma2} it follows that 
\beq
\label{eq:lemma}
	\sum_{j=1}^{m} \| {\bf{a}} - {\bf{a}}^{(i_j)} \|'^{2} \geq 2 {\epsilon}^2 (m-1).
	\eeq

We now define the function 
\beq
\gamma{(x)}=\max \{ 0, 1- \frac{1}{2 \epsilon^2}x^2 \},
\eeq
and for any   ${\bf{a}} \in {\mathbb{E}}^{N}$, we let $\mathcal{M}_{\bf{a}} = \{  {\bf{a}}^{(i_1)} , {\bf{a}}^{(i_2)} \cdots  {\bf{a}}^{(i_m)} \}$ be a subset of $\mathcal{M}_{\epsilon}^{0}$ whose distance from ${\bf{a}}$ is not larger than $\sqrt{2}\epsilon$. We have
\bea
	\sum_{j=1}^{M} \gamma(\| {\bf{a}} - {\bf{a}}^{(j)} \|')
	&=& \sum_{k=1}^{m} \gamma(\| {\bf{a}} - {\bf{a}}^{(i_k)} \|')
	\nonumber \\
	&=& \sum_{k=1}^{m} \left( 1- \frac{1}{2\epsilon^2} \| {\bf{a}} - {\bf{a}}^{(i_k)} \|'^2 \right)
	\nonumber \\
	&=& m- \frac{1}{2\epsilon^2} \sum_{k=1}^{m} \| {\bf{a}} - {\bf{a}}^{(i_k)} \|'^{2}
	\nonumber \\
	&\leq& m-(m-1) \nonumber \\
	&=& 1,
	\label{one}
\eea
where the last inequality follows from (\ref{eq:lemma}). If ${\bf{a}} \notin \mathcal{S}_{\sqrt{E}+\sqrt{2}\epsilon}$, then $\sum_{j=1}^{M} \gamma(\| {\bf{a}} - {\bf{a}}^{(j)} \|')=0$ because $\mathcal{M}_{\bf{a}} = \emptyset$.
By using (\ref{one}) and this last observation, we  perform the following computation:
\bea
	{\rm{vol}} \left( \mathcal{S}_{\sqrt{E}+\sqrt{2}\epsilon} \right)
	&=& \int_{\mathcal{S}_{\sqrt{E}+\sqrt{2}\epsilon}}  d{\bf{ a}}
	\nonumber \\
	&\geq&
	\int_{\mathcal{S}_{\sqrt{E}+\sqrt{2}\epsilon}} \sum_{j=1}^{M} \gamma (\| {\bf{a}} - {\bf{a}}^{(j)}\|') d{\bf{a}}
	\nonumber \\
	&=& \sum_{j=1}^{M} \int_{\mathbb{E}^{N}} \gamma (\| {\bf{a}} - {\bf{a}}^{(j)}\|') d{\bf{a}}
	\nonumber \\
	&=& M \int_{\mathbb{E}^{N}} \gamma (\| {\bf{a}} \|') d{\bf{a}}
	\nonumber \\
	&=& M \int_{0}^{\sqrt{2}\epsilon} \gamma(x) d(\beta_{N}x^{N})
	\nonumber \\
	&=& \beta_{N} M N \int_{0}^{\sqrt{2}\epsilon} \gamma(x) x^{N -1} dx
	\nonumber \\
	&=& \frac{2 \beta_{N}M}{N + 2} (\sqrt{2}\epsilon)^{N},
\eea
where $\beta_{N}$ is  the volume of $\mathcal{S}_{1}$ in $\mathbb{E}^{N}$. Since ${\rm{vol}} \left( \mathcal{S}_{\sqrt{E}+\sqrt{2}\epsilon} \right) = \beta_{N} (\sqrt{E}+\sqrt{2}\epsilon)^{N}$, we obtain
\beq
	M_{\epsilon}^{0} (\BBOP) = M \leq \frac{N +2}{2} \left( 1+\frac{\sqrt{E}}{\sqrt{2}\epsilon} \right)^{N}.
\eeq
The proof is completed by taking the logarithm.
\end{IEEEproof}


\begin{theorem}
\label{thm2}
	For any $0<\delta<1$ and $\epsilon>0$, we have
	\begin{empheq}[left=\empheqlbrace]{align}
		&\bar{C}_{\epsilon}^{\delta}  (\BOP) \geq \frac{\Omega}{\pi} \left[ \log \left(  \frac{\sqrt{E}}{\epsilon} \right) \right], \\
		&\bar{C}_{\epsilon}^{\delta}  (\BOP) \leq \frac{\Omega}{\pi} \left[ \log \left( 1+ \frac{\sqrt{E}}{\epsilon} \right) \right].
	\end{empheq}
\end{theorem}

\begin{IEEEproof}
To prove the result it is enough to show the following inequalities for the $(\epsilon, \delta)$-capacity
\begin{empheq}{align}
	&C_{\epsilon}^{\delta} ( \BOP) \geq N \left[ \log \left( \zeta(N) \frac{\sqrt{E}}{\epsilon} \right) \right] + \log{\delta}, \label{eq:lower2} \\
	&C_{\epsilon}^{\delta} ( \BOP) \leq N \left[ \log \left( 1+ \frac{\sqrt{E}}{\epsilon} \right)
	\right] + \log{\frac{1}{1-\delta}}, \label{eq:upper2}
\end{empheq}
because $\lim_{T \rightarrow \infty} \zeta(N) = 1$ and both $\log{\delta}$ and $\log{\frac{1}{1-\delta}}$ are $o(T)$. 

\noindent
{\emph{Lower bound.}} \;\;
We show that there exists a codebook $\mathcal{M}=\{ {\bf{a}}^{(1)}, {\bf{a}}^{(2)}, \cdots, {\bf{a}}^{(M)} \}$, where
\beq
	M=\delta \left( \zeta(N) \frac{\sqrt{E}}{\epsilon} \right)^{N},
\label{satisfying}
\eeq
 that has cumulative error measure  $\Delta \leq \delta$. To prove this result, we consider an auxiliary stochastic communication model where the transmitter selects a signal uniformly at random from a given codebook and,
given the signal ${\bf{a}}^{(i)}$ is sent, the receiver observes   ${\bf{a}}^{(i)} + {\bf{n}}$, with  ${\bf{n}}$ distributed uniformly in $\mathcal{S}_{\epsilon}$. The receiver compares this signal with all signals  in the codebook and selects the one that is nearest to it  as the one actually sent. 
The decoding error probability of this stochastic communication model, averaged over the uniform selection of signals in the codebook, is given by
\beq
	P_{err} = \frac{1}{M} \sum_{i=1}^{M} \frac{{\rm{vol}}(\mathcal{D}^i)}{{\rm{vol}}(\mathcal{S}^i)},
\label{eq:avd}
\eeq
and by (\ref{di}) and (\ref{cumdi}) it corresponds to the cumulative error measure $\Delta$ of the deterministic model that uses the same codebook. It follows that in order to prove the desired lower bound in the deterministic model, we can  show that there exists a codebook in the stochastic model satisfying (\ref{satisfying}), and whose decoding error probability is at most $\delta$. This follows from a standard  random coding argument, in conjunction to a less standard geometric argument due to the metric employed. 

We construct a random codebook by selecting $M$ signals  uniformly at random inside the ellipsoid $\BOP$. We indicate the average error probability over all signal selections in the codebook and over all codebooks and  by $\bar{P}_{err}$. Since all signals in the  codebook have the same error probability when averaged over all codebooks, $\bar{P}_{err}$ is the same as the average error probability over all codebooks when ${\bf{a}}^{(1)}$ is transmitted. Let in this case the received signal be ${\bf{y}}$ and let $\mathcal{S}_{\epsilon}^{\bf{y}}$ be an Euclidean ball whose radius is $\epsilon$ and center is $\bf{y}$. 

The probability that the signal $\bf{y}$ is decoded correctly is at least as large as the probability that the remaining $M-1$ signals in the codebook are in $\BOP \setminus \mathcal{S}_{\epsilon}^{\bf{y}}$. 
By the union bound, we have
\bea
	1-\bar{P}_{err} 
	&\geq& 1-(M-1) \frac{{\rm{vol}}(\mathcal{S}_{\epsilon}^{\bf{y}})}{{\rm{vol}}(\BOP)}
	\nonumber \\
	&\geq& 1-M \frac{{\rm{vol}}(\mathcal{S}_{\epsilon}^{\bf{y}})}{{\rm{vol}}(\BOP)}	
	\nonumber \\
	&=& 1-M \left( \frac{\epsilon}{\zeta(N) \sqrt{E}} \right)^{N}.
	\label{perr}
\eea
Letting $M=\delta \left( \zeta(N) \frac{\sqrt{E}}{\epsilon} \right)^{N}$, we have $\bar{P}_{err} \leq \delta$. This implies that there exist a given codebook for which the average probability of error over the  selection of signals in the codebook given in  (\ref{eq:avd}) is at most $\delta$. When this same codebook is applied in the deterministic model,  we also have a cumulative error measure $\Delta \leq \delta$.

\noindent
{\emph{Upper bound.}} \;\;
Let $\mathcal{M}_{\epsilon}^{\delta}$ be a maximal $(\epsilon,\delta)$-distinguishable subset of $\BOP$ and $M_{\epsilon}^{\delta} (\BOP)=M$ be the number of elements in $\mathcal{M}_{\epsilon}^{\delta}$. 
Let $\HBOP$ be the union of $\BOP$ and the trace of the inner points of an $\epsilon$-ball whose center is moved along the boundary of $\BOP$, as depicted in Fig.~\ref{fig:slide}. 
\begin{figure}[!t]
\centering
\includegraphics[width=60mm]{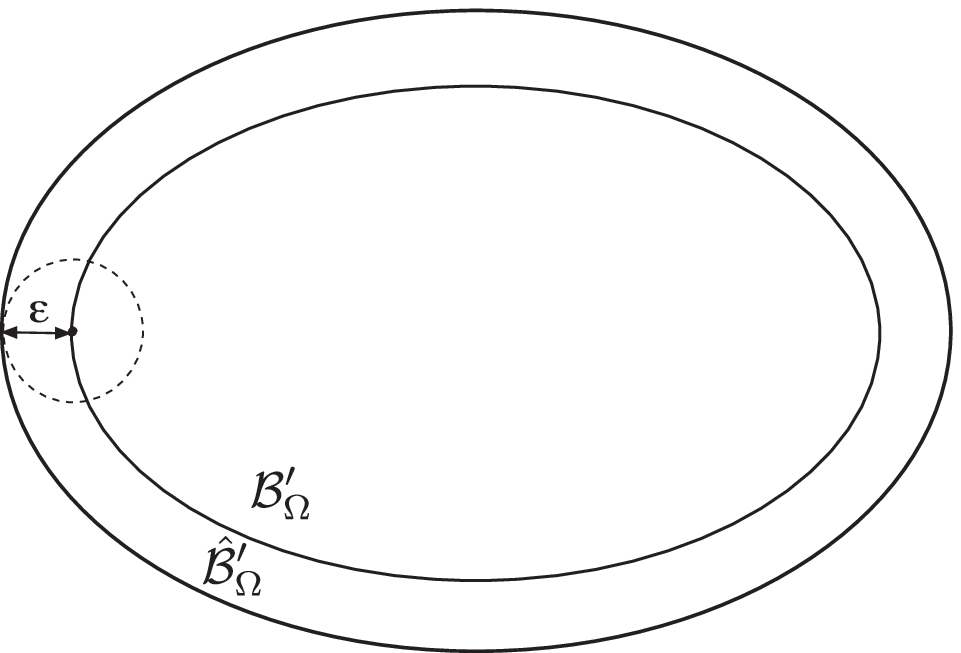}
\caption{Illustration of the relationship between $\BOP$ and $\HBOP$.}
\label{fig:slide}
\end{figure}

Since $\bigcup_{i=1}^{M}{\mathcal{S}}^{i} \subset \HBOP$, we have
\beq
\label{temp}
	{\rm{vol}} \left( \bigcup_{i=1}^{M}{\mathcal{S}}^{i} \right)
	\leq
	{\rm{vol}} \left( \HBOP \right).
\eeq
Since $\bigcup_{i=1}^{M}{\mathcal{S}}^{i} = \biguplus_{i=1}^{M} ({\mathcal{S}}^{i} \setminus {\mathcal{D}}^{i})$, where $\biguplus$ indicates disjoint union, we obtain
\bea
\label{temp2}
	{\rm{vol}} \left( \bigcup_{i=1}^{M}{\mathcal{S}}^{i} \right)
	&=& \sum_{i=1}^{M} \left[ {\rm{vol}}({\mathcal{S}}^{i}) - {\rm{vol}}({\mathcal{D}}^{i}) \right] 
	\nonumber \\
	&=& \sum_{i=1}^{M} {\rm{vol}}({\mathcal{S}}^{i})  \left[ 1- \frac{{\rm{vol}}({\mathcal{D}}^{i})}{{\rm{vol}}({\mathcal{S}}^{i})} \right]
	\nonumber \\
	&=& \sum_{i=1}^{M} {\rm{vol}}({\mathcal{S}}^{i})  \left( 1- \Delta_{i}  \right)
	\nonumber \\
	&=& M \cdot {\rm{vol}}({\mathcal{S}}_{\epsilon}) (1-\Delta).
\eea
Since $\HBOP \subset \mathcal{S}_{\sqrt{E}+\epsilon}$, (\ref{temp}) can be rewritten as
\beq
	M  {\rm{vol}}({\mathcal{S}}_{\epsilon}) (1-\Delta) \leq {\rm{vol}} \left( \mathcal{S}_{\sqrt{E}+\epsilon} \right)
\eeq
or equivalently
\beq
	M_{\epsilon}^{\delta}(\BOP) = M \leq \frac{1}{1-\Delta} \left( 1 + \frac{\sqrt{E}}{\epsilon} \right)^{N}.
\eeq

Since $C_{\epsilon}^{\delta}(\BOP) = \log M_{\epsilon}^{\delta}(\BOP)$ and $\Delta \leq \delta$, the result follows.
\end{IEEEproof}



\begin{theorem}
\label{thm3}
	For any $\epsilon>0$, we have
	\beq
		\bar{H}_{\epsilon} (\BOP) 
		= \frac{\Omega}{\pi} \left[ \log \left( \frac{\sqrt{E}}{\epsilon} \right) \right].
	\eeq
\end{theorem}

\begin{IEEEproof}
To prove the result it is enough to show the following inequalities for the $\epsilon$-entropy
\begin{empheq}{align}
	&H_{\epsilon} ( \BOP) \geq	N \left[ \log \left( \zeta(N) \frac{\sqrt{E}}{\epsilon} \right) \right], \label{eq:lower3} \\
	&H_{\epsilon} ( \BOP) \leq N \left[ \log \left( \frac{\sqrt{E}}{\epsilon} \right) \right] + \eta(N), \label{eq:upper3}
\end{empheq}
where $\eta(N) =o(T)$ and $\lim_{T \rightarrow \infty} \zeta(N) = 1$. 


\noindent {\emph{Lower bound.}} \;\;
Let $\mathcal{L}_{\epsilon}$ be a minimal $\epsilon$-covering subset of $\BOP$ and $L_{\epsilon}(\BOP)$ be the number of elements in $\mathcal{L}_{\epsilon}$. Since $\mathcal{L}_{\epsilon}$ is an $\epsilon$-covering, we have
\beq
\label{eq:c3}
	{\rm{vol}}(\BOP) \leq L_{\epsilon}(\BOP) {\rm{vol}}(\mathcal{S}_{\epsilon}),
\eeq
where $\mathcal{S}_{\epsilon}$ is an Euclidean ball whose radius is $\epsilon$. By combining (\ref{eq:c2}) and (\ref{eq:c3}), we have
\beq
	L_{\epsilon}(\BOP) \geq \left( \zeta(N) \frac{\sqrt{E}}{\epsilon} \right)^N.
\eeq
The proof is completed by taking the logarithm.


\noindent {\emph{Upper bound.}} \;\;
We define the auxiliary set  
\beq
	\BBOP = \left\{ {\bf{a}} = (a_1, a_2, \cdots, a_N) : \sum_{n=1}^{N} a_n^2 \leq E \right\}.
\eeq
The corresponding space $(\BBOP, \|\cdot\|')$ is  Euclidean. Since $\BOP \subset \BBOP$, it follows that $H_{\epsilon}(\BOP) \leq H_{\epsilon}(\BBOP)$ and it is sufficient to derive an upper bound for $H_{\epsilon}(\BBOP)$.

Let $\mathcal{L}_{\epsilon}$ be a minimal $\epsilon$-covering subset of $\BBOP$ and $L_{\epsilon}(\BBOP)$ be the number of elements in $\mathcal{L}_{\epsilon}$. By applying Lemma \ref{lemma3}, we have
\beq
\label{eq:c4}
	L_{\epsilon}(\BBOP) 
	\leq \frac{4e N^{3/2} \left( \frac{\sqrt{E}}{\epsilon} \right)^N }{\ln{N}-2} \left[ N \ln{N} + o(N \ln{N}) \right],
\eeq
for $N \geq 9$ and $1< \frac{\sqrt{E}}{\epsilon} < \frac{N}{\ln{N}}$.   By taking the logarithm, we have
\begin{align}
\label{eq:c5}
	H_{\epsilon}(\BBOP) 
	& \leq  N \log \left( \frac{\sqrt{E}}{\epsilon} \right)  
	\nonumber \\
	& \hspace{.3cm} + \log \left( \frac{4e N^{3/2} }{\ln{N}-2} \left[ N \ln{N} + o(N \ln{N}) \right] \right),
\end{align}
Letting $\eta(N)$ be equal to the second term of (\ref{eq:c5}) the result follows.
\end{IEEEproof}

\subsection{Main theorems for $\BO$}

We now extend results to the full space $\BO$. 
We define the auxiliary set
\beq
	\BBO = \left\{ {\bf{b}} = (b_1, \cdots, b_N, 0, 0, \cdots) : \sum_{n=1}^{N} b_n^2 \leq E \right\}
\eeq
whose norm is the same as   $\BO$. We  also use another auxiliary set
\beq
	\BOPP = \left\{ {\bf{b}} = (b_1, b_2, \cdots, b_{N'}) : \sum_{n=1}^{N'}b_n^2 \leq E \right\},
	\label{hbop}
\eeq
equipped with the norm
\beq
	\|{\bf{b}}\|'' = \sqrt{\sum_{n=1}^{N'} b_n^2 \lambda_n}.
\eeq
where $N' = (1+\alpha)N_0$ for an arbitrary $\alpha>0$.


\begin{theorem}
\label{thm4}
	For any $\epsilon>0$, we have
	\begin{empheq}[left=\empheqlbrace]{align}
		 &\bar{C}_{\epsilon}^{0} (\BO) \geq \frac{\Omega}{\pi} \left[ \log \left(  \frac{\sqrt{E}}{\epsilon} \right) -1 \right]  \label{temp0}\\
		 &\bar{C}_{\epsilon}^{0} (\BO) \leq\frac{\Omega}{\pi} \left[ \log \left( 1+  \frac{\sqrt{E}}{\epsilon \sqrt{2}}  \right) \right].
	\end{empheq}
\end{theorem}

\begin{IEEEproof}
By the continuity of the logarithmic function, to prove the upper bound it is enough to show that for any $\epsilon>\mu>0$
\begin{empheq}{align}
	&\bar{C}_{\epsilon}^{0} (\BO) \leq \frac{\Omega}{\pi} \left[ \log \left( 1+  \frac{\sqrt{E}}{(\epsilon-\mu)\sqrt{2}}  \right) \right],  \label{temp1} 
\end{empheq}
and in order to prove (\ref{temp0}) and (\ref{temp1}),  it is enough to show the following inequalities for the $2\epsilon$-capacity:   for any $\epsilon>\mu>0$
\begin{empheq}{align}
	&C_{\epsilon}^{0} ( \BO) \geq  C_{\epsilon}^{0}(\BOP),	\label{eq:lower4} \\
	&C_{\epsilon}^{0} ( \BO) \leq  C_{\epsilon-\mu}^{0}(\BOP), \label{eq:upper4}
\end{empheq}
and then apply  Theorem~\ref{thm1}.

\noindent {\emph{Lower bound.}} \;\;
Let $\mathcal{D}$ be a maximal $(\epsilon,0)$-distinguishable subset of $\BO$
whose cardinality is $2^{C_{\epsilon}^{0}(\BO)}$.
Similary, let $\mathcal{E}$ be a maximal $(\epsilon, 0)$-distinguishable subset of $\BBO$
whose cardinality is $2^{C_{\epsilon}^{0}(\BBO)}$. 
Note that $\mathcal{E}$ is also a $(\epsilon,0)$-distinguishable subset of $\BO$. Thus, we have
\beq
	2^{C_{\epsilon}^{0}(\BBO)} = |\mathcal{E}| \leq |\mathcal{D}| = 2^{C_{\epsilon}^{0}(\BO)}.
\eeq
From which it follows that
\beq
	C_{\epsilon}^{0}(\BBO) \leq C_{\epsilon}^{0}(\BO).
\eeq 
Since $C_{\epsilon}^{0}(\BBO) = C_{\epsilon}^{0}(\BOP)$, the result follows.


\noindent {\emph{Upper bound.}} \;\;
For any $\epsilon>\mu>0$, we consider a projection map $\beta_{\mu} : \BO \rightarrow \BBO$. 
Let $\mathcal{D}$ be a maximal $(\epsilon,0)$-distinguishable subset of $\BO$
whose cardinality is $2^{C_{\epsilon}^{0}(\BO)}$.
Similary, let $\mathcal{E}$ be a maximal $(\epsilon - \mu,0)$-distinguishable subset of $\BBO$
whose cardinality is $2^{C_{\epsilon - \mu}^{0}(\BBO)}$. 

We define $\mathcal{E'}=\beta_{\mu}(\mathcal{D})$. In general, $\beta_{\mu}$ is not one-to-one correspondence, however $|\mathcal{D}| = |\mathcal{E'}|$. If this is not the case, then there exist a pair of points ${\bf{b}}^{(1)}, {\bf{b}}^{(2)} \in \mathcal{D}$ satisfying 
$\beta_{\mu}({\bf{b}}^{(1)})=\beta_{\mu}({\bf{b}}^{(2)})={\bf{a}}$, and we have
\begin{align}
	\|{\bf{b}}^{(1)} - {\bf{b}}^{(2)} \| 
	&= \| {\bf{b}}^{(1)}-{\bf{a}} + {\bf{a}}-{\bf{b}}^{(2)} \|
	\nonumber \\
	&\leq \|{\bf{b}}^{(1)}-{\bf{a}}\| + \|{\bf{a}}-{\bf{b}}^{(2)}\|
	\nonumber \\
	&\leq \mu + \mu
	\nonumber \\
	&\leq 2\epsilon,
\end{align}
which is a contradiction. Thus, we have
\beq
\label{t1}
	2^{C_{\epsilon}^{0}(\BO)} = |\mathcal{D}| = |\mathcal{E'}|.
\eeq

The distance between any pair of points in $\mathcal{E'}$ exceeds $2(\epsilon - \mu)$.
If this is not the case, then there exist a pair of points in $\mathcal{E'}$ whose distance is smaller than $2(\epsilon - \mu)$. 
These two point can be represented by ${\bf{a}}^{(1)} = \beta_{\mu}({\bf{b}}^{(1)})$ and ${\bf{a}}^{(2)} = \beta_{\mu}({\bf{b}}^{(2)})$, where ${\bf{b}}^{(1)},{\bf{b}}^{(2)} \in \mathcal{D}$. 
It follows that
\begin{align}
	\|{\bf{b}}^{(1)} - {\bf{b}}^{(2)} \| 
	&= \| {\bf{b}}^{(1)}-{\bf{a}}^{(1)}+{\bf{a}}^{(1)}-{\bf{a}}^{(2)}+{\bf{a}}^{(2)}-{\bf{b}}^{(2)} \|
	\nonumber \\
	&\leq \|{\bf{b}}^{(1)}-{\bf{a}}^{(1)}\| + \|{\bf{a}}^{(1)}-{\bf{a}}^{(2)}\|
	\nonumber \\
	& \hspace{.5cm}+\|{\bf{a}}^{(2)}-{\bf{b}}^{(2)}\|
	\nonumber \\
	&\leq \mu + 2(\epsilon - \mu) + \mu
	\nonumber \\
	&\leq 2\epsilon,
\end{align}
which is a contradiction. 
Thus, $\mathcal{E'}$ is a $(\epsilon - \mu,0)$-distingushiable subset of $\BBO$, and we have
\beq
\label{t2}
	|\mathcal{E'}| \leq |\mathcal{E}| = 2^{C_{\epsilon-\mu}^{0}(\BBO)}.
\eeq

By combinining (\ref{t1}) and (\ref{t2}), we obatin
\beq
	2^{C_{\epsilon}^{0}(\BO)} = |\mathcal{D}| = |\mathcal{E'}| \leq |\mathcal{E}| = 2^{C_{\epsilon-\mu}^{0}(\BBO)}. 
\eeq
From which it follows that 
\beq
\label{3-1}
	C_{\epsilon}^{0}(\BO) \leq C_{\epsilon - \mu}^{0}(\BBO).
\eeq
Since $C_{\epsilon-\mu}^{0}(\BBO) = C_{\epsilon-\mu}^{0}(\BOP)$, the result follows.

\end{IEEEproof}


\begin{theorem}
\label{thm5}
	For any $0<\delta<1$ and $\epsilon>0$, we have
	\begin{empheq}[left=\empheqlbrace]{align}
		&\bar{C}_{\epsilon}^{\delta}  (\BO) \geq \frac{\Omega}{\pi} \left[ \log \left(  \frac{\sqrt{E}}{\epsilon} \right) \right] \\
		&\bar{C}_{\epsilon}^{\delta}  (\BO) \leq \frac{\Omega}{\pi} \left[ \log \left( 1+ \frac{\sqrt{E}}{\epsilon} \right) \right].
	\end{empheq}
\end{theorem}

\begin{IEEEproof}
In this case, while the lower bound follows  from a corresponding inequality on the $(\epsilon,\delta)$-capacity, the upper bound follows from an approximation argument and holds for the $(\epsilon,\delta)$-capacity per unit time only.

\noindent {\emph{Lower bound.}} \;\;
Let $\mathcal{E}$ be a maximal $(\epsilon, \delta')$-distinguishable subset of $\BOP$
whose cardinality is $2^{C_{\epsilon}^{\delta'}(\BOP)}$. 
We define a map $\alpha:\BOP \rightarrow \BO$ such that, for ${\bf{b}}=(b_1, \cdots, b_N) \in \BOP$, we have 
\beq
	\alpha({\bf{b}}) = (b_1, \cdots, b_N, 0, 0, \cdots) \in \BO.
\eeq
Then $\alpha(\mathcal{E})$ is a $(\epsilon,\delta'')$-distinguishable subset of $\BO$ where $\delta'' \rightarrow 0$ for  $\delta' \rightarrow 0$. Thus, we can choose $\delta'$ whose corresponding $\delta''$ is smaller than $\delta$. 
In this case, we have
\beq
\label{l5-1}
	2^{C_{\epsilon}^{\delta'}(\BOP)} = |\mathcal{E}| \leq 2^{C_{\epsilon}^{\delta''}(\BO)}.
\eeq 
Also, since $\delta'' < \delta$, we have
\beq
\label{l5-2}
	C_{\epsilon}^{\delta''}(\BO) \leq C_{\epsilon}^{\delta}(\BO).
\eeq
By combining (\ref{l5-1}) and (\ref{l5-2}), we obtain
\beq
	C_{\epsilon}^{\delta'}(\BOP) \leq C_{\epsilon}^{\delta}(\BO).
\eeq
The result now follows from Theorem \ref{thm2}.


\noindent {\emph{Upper bound.}} \;\;
We define 
\beq
	d(\BOPP, \BO) =  \sup_{f \in \BO} \inf_{g \in \BOPP}\|f-g\|
\eeq
which is a measure of distance between $\BOPP$ and $\BO$.
From the Property 6 of the PSWF, we have
\beq
	d (\BOPP, \BO) \rightarrow 0 \;\;\;
	\mbox{as} \;\;\; N_0 \rightarrow \infty.
\eeq
which implies
\beq
	\bar{C}_{\epsilon}^{\delta}  (\BO) = \bar{C}_{\epsilon}^{\delta}  (\BOPP).
\eeq
Thus, in order to prove the upper bound of $\bar{C}_{\epsilon}^{\delta}  (\BO) $, 
it is sufficient to derive an upper bound for $\bar{C}_{\epsilon}^{\delta}  (\BOPP) $.

By using a the same proof technique as the one in Theorem \ref{thm2}, we   obatin
\beq
	C_{\epsilon}^{\delta} ( \BOPP) \leq N' \left[ \log \left( 1+ \frac{\sqrt{E}}{\epsilon} \right)
	\right] + \log{\frac{1}{1-\delta}}
\eeq
which implies
\beq
	\bar{C}_{\epsilon}^{\delta} ( \BOPP) \leq (1+\alpha) \frac{\Omega}{\pi} \left[ \log \left( 1+ \frac{\sqrt{E}}{\epsilon} \right)
	\right].  
\eeq
Since $\alpha$ is an arbitrary positive number, the result follows.

\end{IEEEproof}


\begin{theorem}
\label{thm6}
	For any $\epsilon>0$, we have
	\beq
		 \bar{H}_{\epsilon} (\BO) =	\frac{\Omega}{\pi} \left[ \log \left( \frac{\sqrt{E}}{\epsilon}  \right) \right].
	\eeq
\end{theorem}

\begin{IEEEproof}
By the continuity of the logarithmic function, to prove the result it is enough to show that for any $\epsilon>\mu>0$
\begin{empheq}{align}
	 &\bar{H}_{\epsilon} (\BO) \geq \frac{\Omega}{\pi} \left[ \log \left(  \frac{\sqrt{E}}{\epsilon} \right)  \right], \label{temp3.1} \\
	 &\bar{H}_{\epsilon} (\BO) \leq \frac{\Omega}{\pi} \left[ \log \left( \frac{\sqrt{E}}{\epsilon - \mu}  \right) \right], \label{temp3}
\end{empheq}
and in order to prove (\ref{temp3.1}) and (\ref{temp3}), it is enough to show the following inequalities for the $\epsilon$-entropy: for any $\epsilon>\mu>0$
\begin{empheq}{align}
	&H_{\epsilon} ( \BO) \geq  H_{\epsilon}(\BOP) \label{eq:lower5} \\
	&H_{\epsilon} ( \BO) \leq  H_{\epsilon-\mu}(\BOP),	\label{eq:upper5}	
\end{empheq}
and then apply Theorem~\ref{thm3}.


\noindent {\emph{Lower bound.}} \;\;
For any $\epsilon>\mu>0$, we condiser a projection map $\beta_{\mu} : \BO \rightarrow \BBO$.
Let $\mathcal{D}$ be a minimal $\epsilon$-covering subset of $\BO$
whose cardinality is $2^{H_{\epsilon}(\BO)}$.
Similary, let $\mathcal{E}$ be a minimal $\epsilon$-covering subset of $\BBO$
whose cardinality is $2^{H_{\epsilon}(\BBO)}$. 

We define $\mathcal{E'}=\beta_{\mu}(\mathcal{D})$. 
We  claim that $\mathcal{E'}$ is also a $\epsilon$-covering subset of $\BBO$. 
Let ${\bf{p}}$ be a point of $\BBO$. 
Since $\mathcal{D}$ is an $\epsilon$-covering subset of $\BO$ and $\BBO \subset \BO$, there exists a point ${\bf{b}} \in \mathcal{D}$ such that $\| {\bf{b}} - {\bf{p}} \| \leq \epsilon$.
Note that $\| \beta_{\mu}({\bf{b}}) - {\bf{p}} \| \leq \| {\bf{b}} - {\bf{p}} \|$ and $\beta_{\mu}({\bf{b}}) \in \mathcal{E'}$. This means that, for any point ${\bf{p}} \in \BBO$, there exists a point in $\mathcal{E'}$ whose distance from ${\bf{p}}$ is eqaul or less than $\epsilon$, which implies $\mathcal{E'}$ is a $\epsilon$-covering subset of $\BBO$.
Thus, we have
\beq
\label{s1}
	|\mathcal{E'}| \geq |\mathcal{E}| = 2^{H_{\epsilon}(\BBO)}.
\eeq
Since $|\mathcal{D}| \geq |\mathcal{E'}|$, we obtain the following chain of inequlities:
\beq
	2^{H_{\epsilon}(\BO)} = |\mathcal{D}| \geq |\mathcal{E'}| \geq |\mathcal{E}| = 2^{H_{\epsilon}(\BBO)}.
\eeq
From which it follows that
\beq
	H_{\epsilon}(\BBO) \leq H_{\epsilon}(\BO).
\eeq 
Since $H_{\epsilon}(\BBO) = H_{\epsilon}(\BOP)$, the result follows.


\noindent {\emph{Upper bound.}} \;\;
Let $\mathcal{D}$ be a minimal $\epsilon$-covering subset of $\BO$
whose cardinality is $2^{H_{\epsilon}(\BO)}$.
Similary, let $\mathcal{E}$ be a minimal $(\epsilon-\mu)$-covering subset of $\BBO$
whose cardinality is $2^{H_{\epsilon-\mu}(\BBO)}$. 

We  claim that $\mathcal{E}$ is also an $\epsilon$-covering subset of $\BO$. 
Let ${\bf{p}}$ be a point of $\BO$. Since $\mathcal{E}$ is an $(\epsilon-\mu)$-covering subset of $\BBO$ and $\beta_{\mu}({\bf{p}}) \in \BBO$, there exists a point ${\bf{a}}\in \mathcal{E}$ such that $\| {\bf{a}} - \beta_{\mu}({\bf{p}}) \| \leq \epsilon-\mu$. Then,
\begin{align}
	\| {\bf{a}} - {\bf{p}} \|
	& = \| {\bf{a}} - \beta_{\mu}({\bf{p}}) + \beta_{\mu}({\bf{p}}) - {\bf{p}} \|
	\nonumber \\
	& \leq \| {\bf{a}} - \beta_{\mu}({\bf{p}})\| + \| \beta_{\mu}({\bf{p}}) - {\bf{p}} \|
	\nonumber \\
	& \leq \epsilon - \mu +\mu
	\nonumber \\
	& = \epsilon.
\end{align}
This means that, for any point ${\bf{p}} \in \BO$, there exists a point in $\mathcal{E}$ whose distance from ${\bf{p}}$ is eqaul or less than $\epsilon$, which implies $\mathcal{E}$ is an $\epsilon$-covering subset of $\BO$.
Thus, we have
\beq
	2^{H_{\epsilon-\mu}(\BBO)} = |\mathcal{E}| \geq |\mathcal{D}| = 2^{H_{\epsilon}(\BO)}.
\eeq
From which it follows that
\beq
	H_{\epsilon-\mu}(\BBO) \geq H_{\epsilon}(\BO).
\eeq 
Since $H_{\epsilon-\mu}(\BBO) = H_{\epsilon-\mu}(\BOP)$, the result follows.

\end{IEEEproof}




\appendix

\subsection{Comparison with Jagerman's results}
\label{a:comparison}

A basic relationship between $\epsilon$-entropy and $\epsilon$-capacity, given in~\cite{kolmogorov1961}, is
\begin{equation}{\label{rel}}
	C_{2\epsilon}(\mathcal{A}) \leq H_{\epsilon}(\mathcal{A}).
\end{equation}
It follows that a typical technique to estimate entropy and capacity is to find  a lower bound for $C_{2\epsilon}$ and an upper bound for $H_{\epsilon}$, and if these  are close to each other, then  they are  good estimates for both capacity and entropy.

Following this approach, Jagerman provided a lower bound on the $2\epsilon$-capacity  and an upper bound on the $\epsilon$-entropy of bandlimited functions. In our notation, the lower bound~\cite[Theorem 6]{jagerman1969}  can be written as
\beq
	C_{2\epsilon} \geq N_0 \log \left( \frac{2}{\sqrt{10}} \sqrt{\frac{\SNR_K}{N_0}} +1 \right),
	\label{jag1}
\eeq
where the result is adapted here to real signals. 

\begin{figure}
\begin{center}
{
\scalebox{.85}{\includegraphics{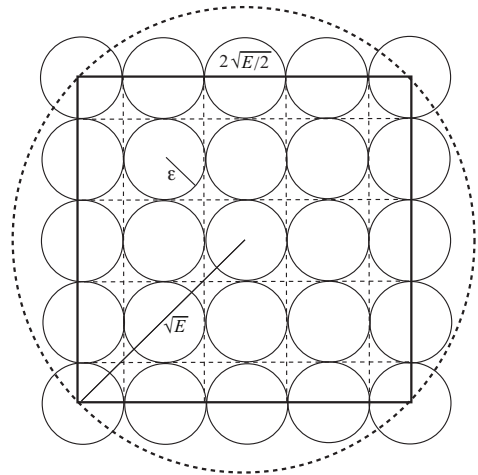}}}
\end{center}
\caption{Lattice packing  in Jagerman's lower bound.}
\label{fig:latticepack}
\end{figure}

Jagerman's  proof roughly follows the  codebook construction corresponding to  the lattice packing depicted in Figure~\ref{fig:latticepack}. 
In higher dimensions the side length of the hypercube corresponding to the  square in Figure~\ref{fig:latticepack} becomes $2\sqrt{E/N_0}$, which divided by the diameter $2 \epsilon$ of the noise sphere  gives the leading term $\sqrt{\SNR_K/N_0}$ inside the logarithm.  The precise result requires a more detailed analysis of the asymptotic dimensionality of the space. This lower bound becomes very loose as $N_0 \rightarrow \infty$. In this case, by using the Taylor expansion of $\log(1+x)$ for $x$ near zero in (\ref{jag1}), it follows that $C_{2\epsilon}$ grows only as $\sqrt{N_0}$ and, as a consequence,  we have the trivial lower bound on the $2 \epsilon$-capacity per unit time
\beq
\label{jager_lower}
	\bar{C}_{2\epsilon} \geq 0.
\eeq
Geometrically, this is due to the volume of the high-dimensional sphere tending to  concentrate on its boundary. For this reason, the packing in the inscribed hypercube in Figure~\ref{fig:latticepack} captures only a vanishing fraction of the volume available in the sphere. 
In contrast, our lower bound in Theorem~\ref{thm1} is non-constructive, and it gives the correct scaling order of the number of bits that can be reliably communicated over the channel, namely $N_0$ rather than $\sqrt{N_0}$, yielding a non-trivial lower bound on the $2 \epsilon$-capacity per unit time.

In the same paper, Jagerman    derives an upper bound on the $\epsilon$-entropy \cite[Theorem 8]{jagerman1969}  by applying  Mitjagin's theorem~\cite{Mitj}, which relates entropy to the Kolmogorov $N$-width. This standard technique is also illustrated in~\cite[Theorem~8]{lorentz}.
For bandlimited signals,  Jagerman further  improves Mityagin's bound in a subsequent paper~\cite[Theorem 1]{jagerman1970}, obtaining in our notation     \beq
\label{jager_upper_temp}
	H_{\epsilon} \leq N \log \left( \frac{2\sqrt{E}}{\epsilon-\mu} + \frac{\epsilon+\mu}{\epsilon-\mu}  \right),
\eeq
where $0 < \mu < \epsilon$ and $N$ is defined in (\ref{eq:choose}). Since $\mu$ is an arbitrary positive number, (\ref{jager_upper_temp}) can be approximated by
\beq
	H_{\epsilon} \leq N  \log \left( 2\sqrt{\SNR_K} + 1 \right).
\eeq
The $\epsilon$-entropy per unit time is then bounded as
\beq
\label{jager_upper}
	\bar{H}_{\epsilon} \leq \frac{\Omega}{\pi} \log ( 2\sqrt{\SNR_K} + 1 ).
\eeq

By combining (\ref{rel}),(\ref{jager_lower}) and (\ref{jager_upper}), Jagerman obtains 
\beq
	0  \leq \bar{H}_{\epsilon} \leq  \frac{\Omega}{\pi} \log \left( 2\sqrt{\SNR_K} + 1 \right),
\eeq
while we  provide a tight characterization of the same quantity in Theorem~\ref{thm6} of this paper.
If we use this tight result to bound the $2\epsilon$-capacity using the classic approach of~(\ref{rel}), we obtain
\beq
\bar{C}_{2 \epsilon} \leq \frac{\Omega}{\pi} \log \sqrt{\SNR_K},
\eeq
while our direct bounds given in Theorem~\ref{thm1} yield, for high values of $\SNR_K$,
\beq
 \frac{\Omega}{\pi} (\log\sqrt{\SNR_K} -1) \leq  \bar{C}_{2 \epsilon} \leq \frac{\Omega}{\pi} (\log \sqrt{\SNR_K}-1/2).
\eeq

\subsection{Relationship with Nair's work}
\label{a:nair}

Nair defined the peak maximum information rate $R_{*}$ in \cite[Lemma 4.2]{nair2013} and showed $R_{*}$ equals the zero-error capacity \cite[Theorem 4.1]{nair2013}. 
In his paper, Nair defined $R_{*}$ for a discrete time channel, but this definition can be modified for a continuous time channel as follows:
\beq
\label{nair01}
	R_{*} = \lim_{T\rightarrow\infty} \sup_{X:X\subset\BO} \frac{I_{*}(X;Y)}{T},
\eeq  
where $Y$ is the uncertain output signal yielded by $X$.

When we consider our channel model, it is clear that the supremum is achieved when $X$ is a maximal $2\epsilon$-distinguishable set, $\mathcal{M}_{2\epsilon}$. In this case, $I_{*}(X;Y) = \log|{X}| = \log{M_{2\epsilon}(\BO)}$.
Thus (\ref{nair01}) can be rewritten as follows:
\beq
\label{nair02}
	R_{*} = \lim_{T\rightarrow\infty} \frac{\log M_{2\epsilon} (\BO)}{T}. 
\eeq

The right-hand side of (\ref{nair02}) is the definition of ${\bar{C}}_{2\epsilon} (\BO)$. Thus, we   conclude that ${\bar{C}}_{2\epsilon} (\BO)$ is a peak maximum information rate and equals the zero-error capacity in our setting.

\subsection{Derivation of the error exponent}
\label{a:err}

By (\ref{perr}), we have
\beq
\Delta =	P_{err} \leq M \left( \frac{\epsilon}{ \zeta(N) \sqrt{E}} \right)^{N}.
\label{above}
\eeq
Let $M=2^{T R}$, where  the transmission rate $R$ is smaller than the lower bound on ${\bar{C}}_{\epsilon}^{\delta}$. Then, (\ref{above}) can be rewritten as
\beq
\Delta	=P_{err} \leq 2^{-T \left[ \frac{N}{T}  \log \left( \zeta(N) \frac{\sqrt{E}}{ \epsilon} \right) -R \right]}.
\eeq
In a stochastic setting the error exponent is defined as the logarithm of the error probability. It follows that we may also define the error exponent in our deterministic model
\beq
	\Er(R) =\frac{N}{T}\log \left( \zeta(N) \frac{\sqrt{E}}{ \epsilon} \right) -R.
\eeq
Since $N/T$ tends to ${\Omega}/{\pi}$ and $\zeta(N)$ tends to $1$ as $T \rightarrow \infty$ , we can approximate the error exponent when $N_0$ is sufficiently large by
\beq
	\Er(R) =\frac{\Omega}{\pi}\log \left( \frac{\sqrt{E}}{\epsilon} \right) -R.
\eeq

\vspace{.5cm}

\noindent  \textbf{Aknowledgments.}
The question of determining a notion of error exponent in a deterministic setting was raised by Francois Baccelli, following the presentation of~\cite{allerton1}.


\bibliographystyle{IEEEtran}
\bibliography{bib_control}

\end{document}